\documentclass[a4paper,12pt]{article} 
\usepackage[square,numbers,sort&compress]{natbib}
\usepackage{epsfig} 
\usepackage{amssymb} 
\usepackage[tbtags]{amsmath}
\usepackage{upref} 
\usepackage[normal]{caption}
\usepackage{float}
\usepackage{wrapfig}
\usepackage{subfigure}
\usepackage{longtable}

\newcommand{\kah}{K\"ahler} 
\newcommand{\nn}{\~n}

\newcommand{\dilaton}{\big(\rule{0pt}{2.1ex}S+\bar S\big)}

\newcommand{\modulit}{\big(\rule{0pt}{2.1ex}T+\bar T\big)}
\newcommand{\moduliz}{\big(\rule{0pt}{2.1ex}Z+\bar Z\big)}

\def\lsim{\raise0.3ex\hbox{$\;<$\kern-0.75em\raise-1.1ex\hbox{$\sim\;$}}}
\def\gsim{\raise0.3ex\hbox{$\;>$\kern-0.75em\raise-1.1ex\hbox{$\sim\;$}}}

\newcommand{\orbifold}{$S^1/Z_2$}

\newcommand{\captions}{\sf\caption}

\def\amu{a_{\mu}}

\def\bsg{$b\to s\gamma$}
\def\asusy{$a_{\mu}^{{\rm  SUSY}}$}
\newcommand{\crosssec}{\sigma_{\tilde\chi^0_1-p}}

\def\preal{{\rm Re\,}}

\def\yzero{\smash{\hbox{$y\kern-4pt\raise1pt\hbox{${}^\circ$}$}}}

\def\s2{\frac{1}{\sqrt2}}

\def\beq{\begin{equation}} 
\def\eeq{\end{equation}} 
\def\beqa{\begin{eqnarray}} 
\def\eeqa{\end{eqnarray}}

\def\IF{\relax{\rm I\kern-.18em F}} 
\def\II{\relax{\rm I\kern-.18em I}} 
\def\IP{\relax{\rm I\kern-.18em P}} 
\def\IC{\relax\hbox{\kern.25em$\inbar\kern-.3em{\rm C}$}} 
\def\IR{\relax{\rm I\kern-.18em R}}

\def\Dsl{\,\raise.15ex\hbox{/}\mkern-13.5mu D} 
\def\IZ{Z\kern-.4em  Z} 
\def\bmat{\left(\begin{array}} 
\def\emat{\end{array}\right)} 
 
\def    \part          {\partial} 
\def    \be            {\begin{equation}} 
\def    \ee            {\end{equation}} 
\def    \bea           {\begin{eqnarray}} 
\def    \eea           {\end{eqnarray}} 
\def    \nn            {\nonumber}

\begin{document} 
\renewcommand{\thefootnote}{\alph{footnote}}

\title{Phenomenology of heterotic M-theory with five-branes
} 
\author{D. G. Cerde\~no\thanks{{\sf dgarcia@delta.ft.uam.es}}, C. Mu\~noz\thanks{{\sf carlos.munnoz@uam.es}}}
\date{}

\thispagestyle{empty}

\rightline{FTUAM 02/16}
\rightline{IFT-UAM/CSIC-02-25}
\rightline{June 2002}
\vspace*{6ex} 

\begin{center}
{\Large\bf Phenomenology of heterotic M-theory with five-branes
}
\\[3ex]

{\large D. G. Cerde\~no\footnote{{\sf dgarcia@delta.ft.uam.es}} and C. Mu\~noz\footnote{{\sf carlos.munnoz@uam.es}}}
\end{center}
\renewcommand{\thefootnote}{\sf\alph{footnote}}
\setcounter{footnote}{0}

\vspace{-3ex}
\begin{center}
{\it 
%
Departamento de F\'\i sica Te\'orica C-XI and 
Instituto de F\'\i sica Te\'orica C-XVI,\\ Universidad Aut\'onoma de Madrid, 
Cantoblanco, 28049 Madrid, Spain.}
\end{center}
\vspace{2ex}

{\abstract{
We analyze some phenomenological implications of heterotic M-theory with five-branes. Recent results for the effective $4$-dimensional action are used to perform a systematic analysis of the parameter space, finding the restrictions that result from requiring the volume of the Calabi-Yau to remain positive. Then the different scales of the theory, namely, the $11$-dimensional Planck mass, the compactification scale and the orbifold scale, are evaluated. The expressions for the soft supersymmetry-breaking terms are computed and discussed in detail for the whole parameter space. 
With this information we study the theoretical predictions for the supersymmetric contribution to the muon anomalous magnetic moment, using the recent experimental result as a constraint on the parameter space.
We finally analyze the neutralino as a dark matter candidate in this construction. In particular, the neutralino-nucleon cross-section is computed and compared with the sensitivities explored by present dark matter detectors.
}}

\allowdisplaybreaks


\clearpage

\section{Introduction}
\setcounter{equation}{0}

The proposal of M-theory as a fundamental theory which contains the five $10$-dimens\-ional superstring theories, as well as $11$-dimensional supergravity, as different vacua of its moduli space has motivated many phenomenological analyses. The cornerstone of most of these works, is the construction due to Ho\v{r}ava and Witten, who showed that the low energy limit of M-theory, compactified on a \orbifold\ orbifold, with $E_8$ gauge multiplets on each of the $10$-dimensional orbifold fixed planes was indeed the strong coupling limit of the $E_8\times E_8$ heterotic string theory \cite{hw95-1,hw96-1}.

A resulting $4$-dimensional $N=1$ supergravity can be obtained if the six remaining extra dimensions are compactified on a Calabi-Yau manifold \cite{witten96-1}. 
This construction possesses a certain number of phenomenological virtues.
The most relevant one is the possibility of tuning the $11$-dimensional Planck scale and the orbifold radius so that the Planck scale, $M_{Planck}=1.2\times10^{19}$ GeV, and the GUT scale, $M_{GUT}\approx3\times10^{16}$ GeV, which is here identified with the inverse of the Calabi-Yau volume, are recovered \cite{witten96-1,bd96-1}.
In the context of this so called heterotic M-theory, the construction with standard embedding for the spin connection into the gauge fields has been thoroughly investigated.  
However, although in the weakly coupling limit of heterotic string, the calculations under the assumption of non-standard embedding  
are more complicated than in the standard embedding cases,  
this is not the case in the strong coupling limit, as emphasized in \cite{low99-1}. 
In this context, non-perturbative objects of M-theory, such as M$5$-branes, can be shown to survive the orbifold projection of Ho\v{r}ava-Witten construction under certain circumstances, permitting much more freedom to play with gauge groups and with the matter fields that appear \cite{dlow98-1,dlow99-1,dow99-1,dopw99-1}. In addition interesting Yukawa textures may arise \cite{ad00-1}. 
The analysis of the resulting nonperturbative vacua studying the gauge kinetic functions and \kah\ potential was first performed in \cite{low98-4,low99-1} and completed in \cite{ds00-3,mps00-1}. In particular, in the latter, the modulus of the five-brane was correctly identified. Also, the effect of the five-brane in the kinetic terms of the \kah\ potential was evaluated by different methods in \cite{ds00-3} and \cite{mps00-1}, and confirmed in \cite{bl01-1}.

These results make it possible to complete former phenomenological analyses \cite{cm99-1,kks99-1}. In particular, the scales of the theory and the new structure of soft supersymmetry-breaking terms can be determined, analyzing the effect of the five-branes on both.
Using these soft parameters, and knowing the initial scale for their running, the low energy supersymmetric spectrum can be obtained, making it possible to extract predictions for low-energy observables.
For example, the theoretical predictions for the supersymmetric contribution to the muon anomalous magnetic moment can be calculated and compared with the recent measurement in the E821 experiment at the BNL. 
Thus we can derive interesting constraints on the parameter space. Also, there has been recently some theoretical activity analyzing the compatibility of regions in the parameter space of supersymmetric theories with the sensitivity of current dark matter detectors. 
In this sense, we can evaluate the neutralino-nucleon cross-section taking into account the different experimental constraints.
We will carry out this analysis in the present work.

In Sec.~\ref{sec_mtheory} the effective supergravity obtained from heterotic M-theory with five-branes is reviewed. We will concentrate on the recently computed corrections on the \kah\ potential and gauge kinetic functions due to the inclusion of a five-brane, and the identification of the correct modulus. 
We will then analyze the parameter space of the theory. Requiring the volume of the Calabi-Yau to remain positive, we will derive the corresponding constraints for the different regions in the parameter space. 
The structure of the scales of the construction will be analyzed in Sec.~\ref{sec_scales}. Different possibilities arise for these scales and we will find that lowering their values is possible in some special limits. Nevertheless, either a fine tuning of the parameters or a large hierarchy between them is needed, rendering this possibility unnatural.
The new expressions of the soft supersymmetry-breaking terms, including the corrections due to the five-brane are computed in Sec.~\ref{sec_softterms}, and their structure is analyzed for representative cases of the parameter space.
We find that an interesting pattern of soft terms arises. In particular, scalar masses larger than gaugino masses are obtained more easily than in the case without five-branes. This is shown with specific examples for several special limits of the parameter space. 
Using these results, we undertake the analysis of low-energy observables. In particular, in Sec.~\ref{sec_muon} the theoretical predictions for the muon anomalous magnetic moment are evaluated and the results compared with recent experimental results.
Finally, in Sec.~\ref{sec_darkmatter} the dark matter implications of this construction are investigated with the evaluation of the neutralino-nucleon cross-section.


\section{M-Theory on \orbifold$\times CY_3$ with five-branes}
\setcounter{equation}{0}
\label{sec_mtheory}

The solution to the equations of motion of $11$-dimensional M-theory \cite{witten96-1,hw96-1} compactified on
\begin{equation}
M_4\times S^1/Z_2\times CY_3\ ,
\end{equation}
were $CY_3$ is a $6$-dimensional Calabi-Yau manifold and $M_4$ is the $4$-dimensional Minkowski space, can be analyzed by an expansion in powers of the dimensionless parameter \cite{bd96-1}
\begin{equation}
\epsilon_1=\frac{\pi\rho}{M^3_{11}V^{2/3}}\ ,
\end{equation}
where $M_{11}$ denotes the $11$-dimensional Planck mass, $V$ is the volume of the Calabi-Yau and $\pi\rho$ denotes the length of the $11$th segment.
The resulting effective $4$-dimensional supergravity was computed to leading order in \cite{bd96-1,lln97-2,dg97-1,noy97-1}. The order $\epsilon_1$ correction to the leading order gauge kinetic functions and \kah\ potential was computed in \cite{bd96-1,choi97-1,ns97-1,noy97-1}, and \cite{low97-1}, respectively. 

It was further investigated if $M5$-branes \cite{witten96-1} of M-theory survived the action of the orbifold and whether they respected or not the same supersymmetries. It can be shown that this is the case if they are parallel to the orbifold fixed hyperplanes \cite{llo97-1}. Moreover, in order to keep $4$-dimensional Lorentz invariance, these have to expand the uncompactified $4$-dimensional space. As a last condition, it can be seen that keeping $N=1$ supersymmetry in four dimensions is only possible if the brane wraps a holomorphic $2$-cycle on the Calabi-Yau.

\subsection{$4$-dimensional effective action}

After compactification, we are left with some chiral superfields which constitute the moduli of the theory. These are the two model independent bulk superfields, the dilaton, $S$, and the modulus, $T$, and a modulus, $Z$, parameterizing the five-brane position along the orbifold\footnote{
We are assuming compactification on a Calabi-Yau manifold with only one \kah\ modulus $T$ (also valid in the overall modulus case), which leads to interesting phenomenological virtues as emphasized in \cite{low99-1}.  In particular,
the soft supersymmetry-breaking terms are automatically universal,
and therefore the presence of dangerous flavour changing neutral currents 
is avoided. Examples of such compactifications exist, as e.g.
the quintic hypersurface $CP^4$. Although these spaces were also
known in the context of the weakly-coupled heterotic string,
the novel fact in heterotic M-theory is that model building 
is relatively simple. For example, in the presence of 
non-standard embedding and five-branes the construction of
three generation models might be considerably easier than for
the standard embedding. 
On the other hand, we are assuming that there is only one modulus $Z$, corresponding to the existence of a single five-brane. Scenarios with more five-branes are possible but would yield qualitatively similar results in the phenomenological analyses performed in this paper.
}. Their scalar components can be expressed in the following way \cite{ds00-3,mps00-1}:
\begin{subequations}
\label{moduli}
\begin{align}
\dilaton&=\frac1{\pi(4\pi)^{2/3}}M_{11}^6V + \modulit\beta_1z^2\ ,\\
\modulit&=\frac{6^{1/3}}{(4\pi)^{4/3}}M_{11}^3V^{1/3}\pi\rho ,\\
\moduliz&=\rule{0pt}{3.1ex}\modulit\beta_1\,z\ \label{moduliz},
\end{align}
\end{subequations}
where $z\in(0,1)$ is the normalized position of the five-brane along the eleventh dimension\footnote{When the five-brane coincides with one of the orbifold fixed planes (i.e. $z\rightarrow0$ or $z\rightarrow1$) new massless states would appear, originated from membranes stretched between the five-brane and the boundary or the five-brane and its $Z_2$ mirror. The theory then undergoes a small-instanton transition \cite{witten95-4,gh96-1,opp00-1,cgl01-1,bdo02-1} and the particle content and other properties of the $N=1$ theory on the relevant boundary change substantially. The dynamics are not well understood at present, and therefore we will not consider here such critic conditions.\label{foot_critic}}, and $\beta_1$ is the charge of the five-brane, which has to be positive in order to preserve supersymmetry \cite{low98-4}.
The resulting $4$-dimensional effective supergravity is then expressed in terms of the \kah\ function\footnote{Here, and in the rest of the article the subscript 'O' stands for 'observable' and 'H' for 'hidden', referring to the fixed hyperplanes of the orbifold.} \cite{ds00-3,mps00-1}:
\begin{align}
K&=-\ln \left(\dilaton-{\frac {{\moduliz}^{2}}{\beta_1\,\modulit}}\right)-3\,\ln \modulit\ \nonumber\\
&\quad\,+\frac3\modulit\left(1+\frac{e_O}{3}\right)H_{IJ}C^I_O\bar C^{\bar J}_O
+\frac3\modulit\left(1+\frac{e_H}{3}\right)H_{IJ}C^I_H\bar C^{\bar J}_H\ ,
\label{kahlerfunction}
\end{align}
and the gauge kinetic functions:
\begin{subequations}
\label{gaugekinetic}
\begin{align}
f_O&=S+\left(\beta_O+\beta_1\right)T-2Z\ ,\\
f_H&=S-\left(\beta_O+\beta_1\right)T\ .
\end{align}
\end{subequations}
In the expressions above $H_{IJ}$ is a moduli-independent matrix. The integer model-dependent quantities $\beta_O$ and $\beta_H$ are the instanton number on each of the fixed planes, defined as $\beta_{O,H}=\frac{1}{8\pi^2}\int\omega\wedge[{\rm tr}(F_{O,H}\wedge F_{O,H})-\frac12 {\rm tr}(R\wedge R)]$. Together with the five-brane charge, they satisfy
\begin{equation}
\beta_O+\beta_1+\beta_H=0\ ,
\end{equation}
Also, we have defined:
\begin{equation}
e_O=b_O\frac\modulit\dilaton \quad \ ,\quad e_H=b_H\frac\modulit\dilaton\ ,
\label{eoeh}
\end{equation}
where
\begin{eqnarray}
b_O=\beta_O+\beta_1\left(1-z\right)^2\ \quad,\quad\quad
b_H=\beta_H+\beta_1\left(z\right)^2\ ,
\label{bobh}
\end{eqnarray}
and the position of the five-brane is expressed from (\ref{moduliz}) in terms of the moduli as
\begin{equation}
z=\frac\moduliz{\beta_1\,\modulit}\ .
\end{equation}


\subsection{Constraints on the parameter space.}
\label{sec_constraints}

The real parts of the gauge kinetic functions (\ref{gaugekinetic}) are related to the coupling constants on the observable and hidden hyperplanes ($\alpha_O$ and $\alpha_H$ respectively) as $4\pi\, \preal f_{O,H}=1/\alpha_{O,H}$. Thus we can write, using expressions (\ref{gaugekinetic}), (\ref{eoeh}), (\ref{moduli}) and (\ref{bobh}) 
\begin{align}
\alpha_O
&=\frac{1}{2\pi\dilaton\left(1+e_O\frac{\beta_O+\beta_1(1-2z)}{b_O}\right)}\nn\\
&=\frac{\left(4\pi\right)^{2/3}}{2M_{11}^6V_O}\ ,\label{alphao}\\
\alpha_H
&=\frac{1}{2\pi\dilaton\left(1+e_O\frac{\beta_H}{b_O}\right)}\nn\\
&=\frac{\left(4\pi\right)^{2/3}}{2M_{11}^6V_H}\ ,\label{alphah}
\end{align}
where the volume of the Calabi-Yau on the observable and hidden planes is\footnote{
These expressions correspond to a first order approximation in $k^{2/3}$ (where $k^2=M_{11}^{-9}$), in order to be consistent with the precision with which the \kah\ potential is known. For the full dependence of the Calabi-Yau volume on the orbifold coordinate in the context of warped metrics, taking also into account the effect of five-branes, see \cite{ck00-1}.
} :
\begin{alignat}{1}
V_O&=V(1+\sigma e_O)\ ,\label{volume-obs}\\
V_H&=V(1+\sigma e_H)\ ,\label{volume-hid}
\end{alignat}
with
\begin{equation}
\sigma\equiv\frac{\dilaton}{\dilaton-\modulit\beta_1z^2}=\frac{1}{1-\frac{e_O}{b_O}\beta_1z^2}\ge1\ .
\label{sigmadef}
\end{equation}
If we assume that the gauge group of the observable sector is the standard model one or some unification gauge group and we want to reproduce the LEP data about gauge coupling unification, $\alpha_O\approx1/24$, we see that from (\ref{alphao}) we can obtain the useful relation:
\begin{equation}
e_O=\frac{b_O}{\beta_O+\beta_1(1-2z)}
\left(\frac{4-\dilaton}{\rule{0pt}{2.1ex}S+\bar S}\right)\ .
\label{eos}
\end{equation}
Thus we can work with the parameter $e_O$ and recover the values of the different moduli by using (\ref{eos}), (\ref{eoeh}) and (\ref{moduliz}).

Up to now, the parameter space of the theory is determined by four free parameters, which can be chosen as follows: the five-brane position and charge ($z$ and $\beta_1$), the instanton number of the observable hyperplane ($\beta_O$), and the parameter $e_O$ (or $\sigma e_O$). Although $e_O$ ($\sigma e_O$) is a free parameter, its range of values is constrained in order to make sure that the Calabi-Yau volume remains positive
at every point of the orbifold. We will derive here these constraints.

\begin{figure}[!t]
\begin{center}
\epsfig{file=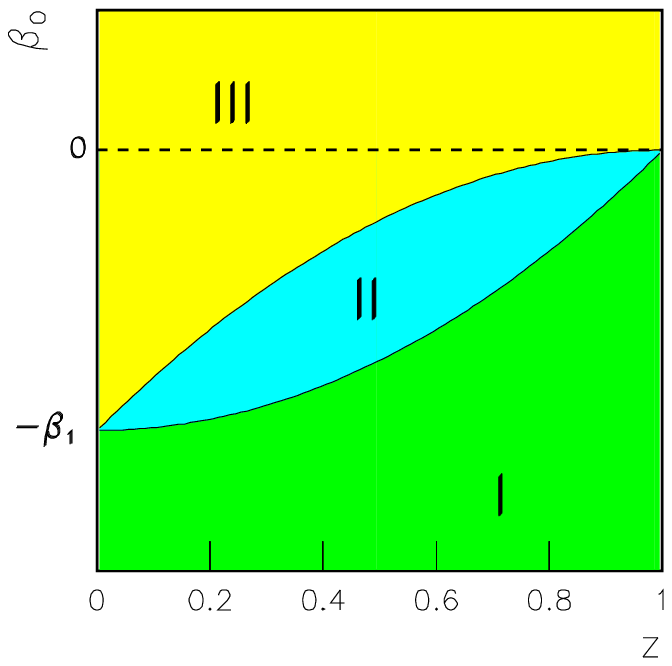,height=7cm}
\end{center}
\vspace{-1cm}
\captions{Different regions of the parameter space discussed in the text
\label{regionsfig}}
\end{figure}
Due to the linear dependence of the volume of the Calabi-Yau on the orbifold direction we only need to impose $V_{O,H}\ge 0$. 
Notice in this sense that to have a positive volume in the observable sector is not sufficient since the volume in the hidden sector might be negative for some choices of the parameters.
Using then (\ref{volume-obs}), (\ref{volume-hid}) and (\ref{eoeh}) we can summarize these constraints as follows:
\begin{subequations}
\label{constraintss}
\begin{eqnarray}
{\sf (I)}\ \ \ b_H\ge 0,\ b_O\le0	&\rightarrow&-1<\sigma e_O\le 0\ ,
\label{constraintssi}\\
{\sf (II)}\ \ b_H< 0,\ b_O\le0		&\rightarrow&\max(-1,-\frac{b_O}{b_H})<\sigma e_O\le 0\ ,\label{constraintssii}\\
{\sf (III)}\ b_H< 0,\ b_O>0		&\rightarrow&0<\sigma e_O<\frac{b_O}{|b_H|}\label{constraintssiii}\ .
\end{eqnarray}
\end{subequations}
The expressions above show no constraints for the case $b_H\ge 0,\ b_O>0$. It can be seen from (\ref{bobh}) that such values cannot be obtained for positive $\beta_1$. Also, from (\ref{bobh}) we find that $\beta_1\ge 0$ implies $b_O\le|b_H|$, and therefore $0\le\sigma e_O< 1$ in (\ref{constraintssiii}).

These constraints can be expressed in a more adequate way in terms of the charge and position of the five-brane and the instanton number of the observable hyperplane. For example, from (\ref{bobh}) it can be seen that (\ref{constraintssi})
implies $\beta_O\le-\beta_1(1-z^2)$. We show this in Fig.\,\ref{regionsfig} as area {\sf (I)}. Following the same arguments, we find that (\ref{constraintssii}) implies $-\beta_1(1-z^2)<\beta_O\le-\beta_1(1-z)^2$, and (\ref{constraintssiii}) implies $-\beta_1(1-z)^2<\beta_O$. The corresponding regions for both cases are also shown in Fig.\,\ref{regionsfig} as areas {\sf (II)} and {\sf (III)}, respectively.

Notice however that $\sigma$ has a dependence on $e_O$, as it can be seen in (\ref{sigmadef}). Using this expression on (\ref{constraintss}) these constraints can be expressed as constraints on $e_O$ as
\begin{subequations}
\label{constraintse}
\begin{eqnarray}
&&{\sf (I)}\ \ \ -\frac{1}{1-\frac{\beta_1}{b_O}z^2}<e_O\le 0\ ,
\label{constraintsei}\\
&&{\sf (II)}\ \ \max(-\frac{1}{1-\frac{\beta_1}{b_O}z^2},-\frac{b_O}{b_H}\frac{1}{1-\frac{\beta_1}{b_H}z^2})<e_O\le 0\ ,\label{constraintseii}\\
&&{\sf (III)}\ 0<e_O<\frac{b_O}{|b_H|}\frac{1}{1+\frac{\beta_1}{|b_H|}z^2}\label{constraintseiii}\ .
\end{eqnarray}
\end{subequations}

\begin{figure}[!t]
\begin{center}
\epsfig{file=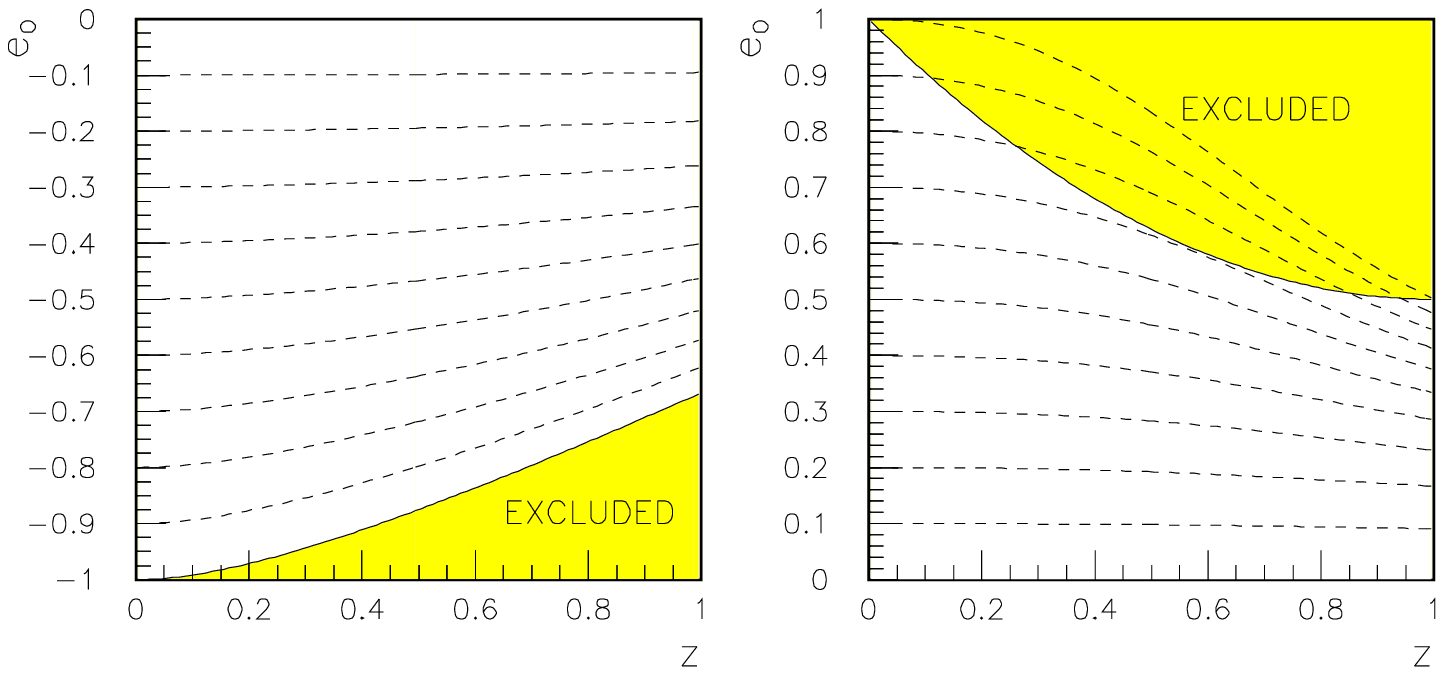,height=7cm}
\end{center}
\vspace{-1cm}
\captions{Allowed (white) and excluded (shaded) areas in the parameter space $e_O - z$ for region {\sf (I)} on the left (for $\beta_O=-2$ and $\beta_1=1$), and region {\sf (III)} on the right (for $\beta_O=1$ and $\beta_1=1$).
\label{exclusion}}
\end{figure}
We illustrate these constraints with some specific examples. In particular, if we take $\beta_O=-2$ and $\beta_1=1$, we can see in Fig.\,\ref{regionsfig} that this corresponds to case {\sf(I)} for every value of the position of the five-brane, $z$. According to (\ref{constraintsei}), the constraints on the parameter $e_O$ will depend on $z$. This is shown on the left-hand side of Fig.\,\ref{exclusion}, where the allowed values for $e_O$ lie in the white area and the shaded area is excluded by these constraints. 
Similarly, the right-hand side of Fig.\,\ref{exclusion} shows an example of the excluded regions for the case $\beta_O=1$ and $\beta_1=1$, which according to Fig.\,\ref{regionsfig}, corresponds to positive values of $e_O$ in region {\sf (III)}. 
Now the area excluded by constraint (\ref{constraintseiii}) is shown as the shaded area. 
For comparison, in both graphs we display as dashed lines the lines along which $\sigma e_O$ is kept constant. It can be seen how the permitted values for $\sigma e_O$ have different constraints, corresponding to (\ref{constraintssi}) and (\ref{constraintssiii}).

\begin{figure}[!t]
\begin{center}
\epsfig{file=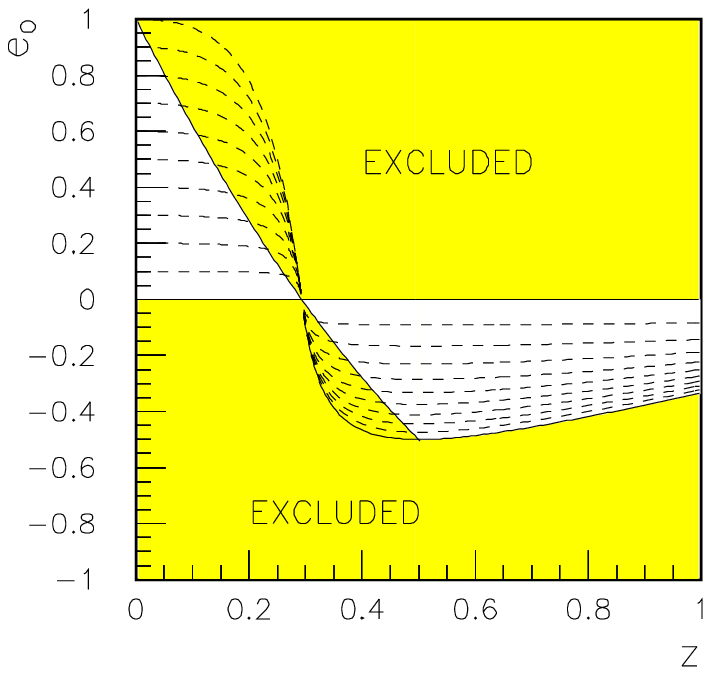,height=7cm}
\end{center}
\vspace{-1cm}
\captions{Allowed area (white) in the parameter space $e_O - z$ for a particular case where region {\sf (II)} is crossed. We have chosen $\beta_O=-1$; $\beta_1=2$ (see Fig.\,\ref{regionsfig}).
\label{exclusionii}}
\end{figure}

In Fig.\,\ref{regionsfig} we also see how if $-\beta_1<\beta_O<0$, varying the position of the five-brane we can move along the different regions in the parameter space. In particular, let us consider the case $\beta_O=-1$ and $\beta_1=2$. From Fig.\,\ref{regionsfig} we see that if the five-brane is close to the observable hyperplane ($z\approx0$), then the constraints corresponding to region {\sf (III)} (\ref{constraintseiii}) must apply, and therefore, we have positive values of $e_O$. If the five-brane moves towards the hidden fixed hyperplane, it arrives at a point which separates regions {\sf (III)} and {\sf (II)} where $b_O=0$. In our case, this point corresponds to $z\approx0.29$. From this point, only negative values of $e_O$ will be allowed, as (\ref{constraintseii}) shows. Finally, $z\approx0.71$ separates regions {\sf (II)} and {\sf (I)}. All these features, together with the corresponding excluded regions due to the different constraints on $e_O$, depending on the region, are shown in Fig.\,\ref{exclusionii}.

The case without five-branes is recovered for $\beta_1=0$. In that case, we have from (\ref{bobh}) $b_O=\beta_O=-b_H$, and from (\ref{sigmadef}) $\sigma=1$. Thus, region {\sf (II)} degenerates into the line $\beta_O=0$ and we are only left with region {\sf (I)}, which now describes the non-standard embedding case with the constraint $-1<e_O\le0$, and region {\sf (III)}, which describes the standard embedding with the constraint $0<e_O<1$, as can be seen from Fig.\,\ref{regionsfig} and expressions (\ref{constraintse}).

Let us describe now some specific scenarios which will be used along this article. The parameters $\beta_O$, $\beta_1$, and $z$ will be fixed and the corresponding constraints on $\sigma e_O$ and $e_O$ can be computed by using the information of Fig.\,\ref{regionsfig}, (\ref{bobh}), (\ref{constraintss}) and (\ref{constraintse}).
\begin{eqnarray}
\beta_O=-2\ ;\ \beta_1=1\ ;\ z=0.50&\ \rightarrow\ &-1<\sigma e_O\le0\ ;\ -\frac78< e_O\le0\ ,\label{scenarioi}\\
\beta_O=-1\ ;\ \beta_1=2\ ;\ z\approx0.29&\ \rightarrow\ &\sigma e_O=e_O=0\ ,\label{scenarioii}\\
\beta_O=\hphantom{-}1\ ;\ \beta_1=1\ ;\ z=0.50&\ \rightarrow\ &0<\sigma e_O<\frac57\ ;\ 0<e_O<\frac58\ .\label{scenarioiii}
\end{eqnarray}
Scenarios (\ref{scenarioi}) and (\ref{scenarioiii}) correspond to regions {\sf(I)} and {\sf(III)}, respectively. The position of the five-brane in (\ref{scenarioii}) is tuned to obtain $b_O=0$ (and therefore $e_O=0$), and corresponds to the case described in Fig.\,\ref{exclusionii}.

Let us finally remark that we take in our analysis the values of the moduli as free parameters. For different attempts to determine these dynamically see \cite{ckk98-1,ck01-1}.


\section{Scales}
\setcounter{equation}{0}
\label{sec_scales}

Heterotic M-theory can reconcile the observed Planck scale, $M_{Planck}=1.2\times10^{19}$ GeV, with the phenomenologically favoured GUT scale, $M_{GUT}\approx3\times10^{16}$ GeV, in a natural manner \cite{witten96-1,bd96-1}. This is still true if higher order corrections are taken into account \cite{cm99-1}. However, the effect that introducing five-branes has on the scales must now be revisited due to the changes we have just described.

Using the expression (\ref{volume-obs}) of the volume of the Calabi-Yau in the observable hyperplane, the M-theory expression for the $4$-dimensional Planck scale
\begin{equation}
M^2_{Planck}=16\pi^2\rho M_{11}^9\langle V\rangle\ ,
\label{planckmass}
\end{equation}
where $\langle V\rangle$ stands for the average volume of the Calabi-Yau ($(V_O+V_H)/2$), and the definitions of the scalar components of the moduli fields (\ref{moduli}) we can derive an expression for the compactification scale $V_O^{-1/6}$
\begin{eqnarray}
V_O^{-1/6}&=&\left(\frac{V}{\langle V\rangle}\right)^{1/2}3.6\times10^{16}
\left(\frac{4}{\dilaton-\modulit\beta^1z^2}\right)^{1/2}\nonumber \\
&&\left(\frac{2}{\rule{0pt}{2.1ex}T+\bar T}\right)^{1/2}
\left(\frac{1}{1+\sigma e_O}\right)^{1/6}\ {\rm GeV}\ .
\label{scales_volume}
\end{eqnarray}
 The rest of the scales of the theory, namely, the $11$-dimensional Planck mass, $M_{11}$, and the orbifold scale, $(\pi\rho)^{-1}$, can be related to $V_O^{-1/6}$ if we use (\ref{planckmass}) and (\ref{alphao}). Also, (\ref{scales_volume}) can be simplified by using (\ref{sigmadef}), (\ref{eos}) and assuming $b_O\ne0$.
The resulting expressions read:
\begin{subequations}
\label{scales-sigmae}
\begin{align}
V_O^{-1/6}&=3.6\,\cdot10^{16}\left(\frac1{1+\frac{\sigma e_O}2\left(1+\frac{b_H}{b_O}\right)}\right)^{1/2}\left(\frac {b_O}{2\sigma e_O}\right)^{1/2}\left(1+\sigma e_O\right)^{5/6}\ {\rm GeV}\ ,\\
\frac{M_{11}}{V_O^{-1/6}}&=\left(2(4\pi)^{-2/3}\alpha_O\right)^{-1/6}\approx 2\ ,\label{scales-sigmaem}\\
\frac{V_O^{-1/6}}{(\pi\rho)^{-1}}&=
\left(
\frac1{1+\frac{\sigma e_O}2\left(1+\frac{b_H}{b_O}\right)}\right)
\left(\frac{2.7\times10^{16}\,{\rm GeV}}{V_O^{-1/6}}\right)^2\left(8192\pi^4\alpha^3\right)^{1/2}(1+\sigma e_O)\ ,
\end{align}
\end{subequations}
where $\alpha_0\approx1/24$ has been used in (\ref{scales-sigmaem}). 
Note that these expressions do not differ from the ones obtained in \cite{cm99-1}, but using $\sigma e_O$ instead of $e_O$. This is because there have been no corrections to the expressions of the gauge kinetic functions (\ref{gaugekinetic}), when expressed in terms of the scales of the theory, which can be done by using (\ref{moduli}) on these. 
We must, however, be aware of the constraints on $\sigma e_O$, summarized in (\ref{constraintss}).

\begin{figure}[!t]
\begin{center}
\epsfig{file=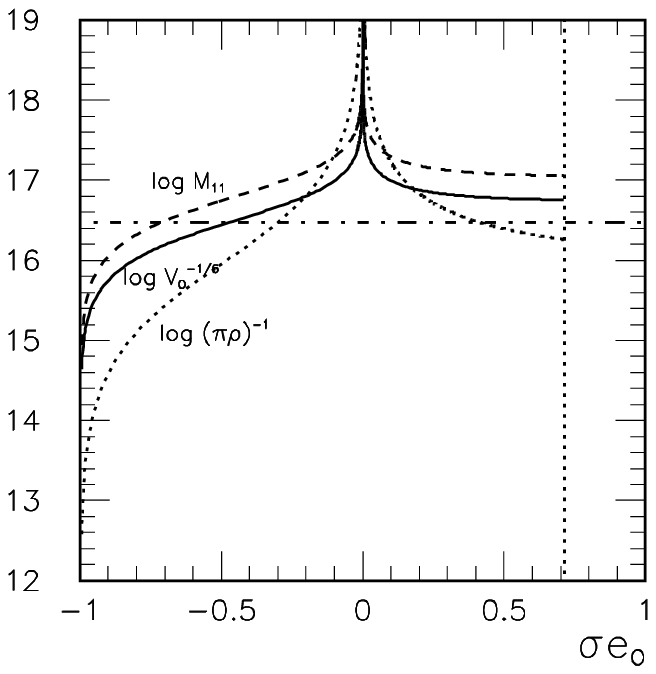}
\end{center}
\vspace*{-1cm}
\captions{Scales of the theory versus $\sigma e_O$ for scenarios (\ref{scenarioiii}) and (\ref{scenarioi}), allowing for positive and negative $e_O$, respectively. The phenomenologically favoured value for the GUT scale, $M_{GUT}=3\times10^{16}$ GeV, is shown as a dot-dashed line.
\label{scalesancient}}
\end{figure}

Let us now consider some representative examples. First, for negative values of $\sigma e_O$ the scenario described in (\ref{scenarioi}) will be used, where $-1<\sigma e_O<0$. The resulting values for the different scales are shown in the left hand side of Fig.\,\ref{scalesancient}. We can see how the phenomenologically favoured value of the GUT scale is easily reached. In particular, in this case, which is the same one of Fig.\,$6$a of \cite{cm99-1}, this happens for $\sigma e_O\approx-0.46$ (i.e. $e_O\approx-0.43$, by (\ref{sigmadef})), which, using (\ref{eos}), (\ref{eoeh}) and (\ref{moduliz}), corresponds to $\dilaton\approx7.9$, 
$\modulit\approx1.95$, 
and $\moduliz\approx0.97$. As it was explained in \cite{cm99-1}, there is now the possibility of lowering the scales when $\sigma e_O\rightarrow-1$ at the price of introducing a fine-tuning problem between the VEVs of the different moduli. We refer to \cite{cm99-1}, where this case is explained in detail. In fact, these results are qualitatively equal to what one obtains in the case without five-branes. This can be found in Fig.\,2 of \cite{cm99-1}.
Likewise, the $e_O\rightarrow 0$ case, where both the radius of the orbifold and the volume of the Calabi-Yau become very small is analogous to the situation discussed in \cite{cm99-1}, where it is argued that this corresponds to the weakly coupled limit.

\begin{figure}[!t]
\begin{center}
\epsfig{file=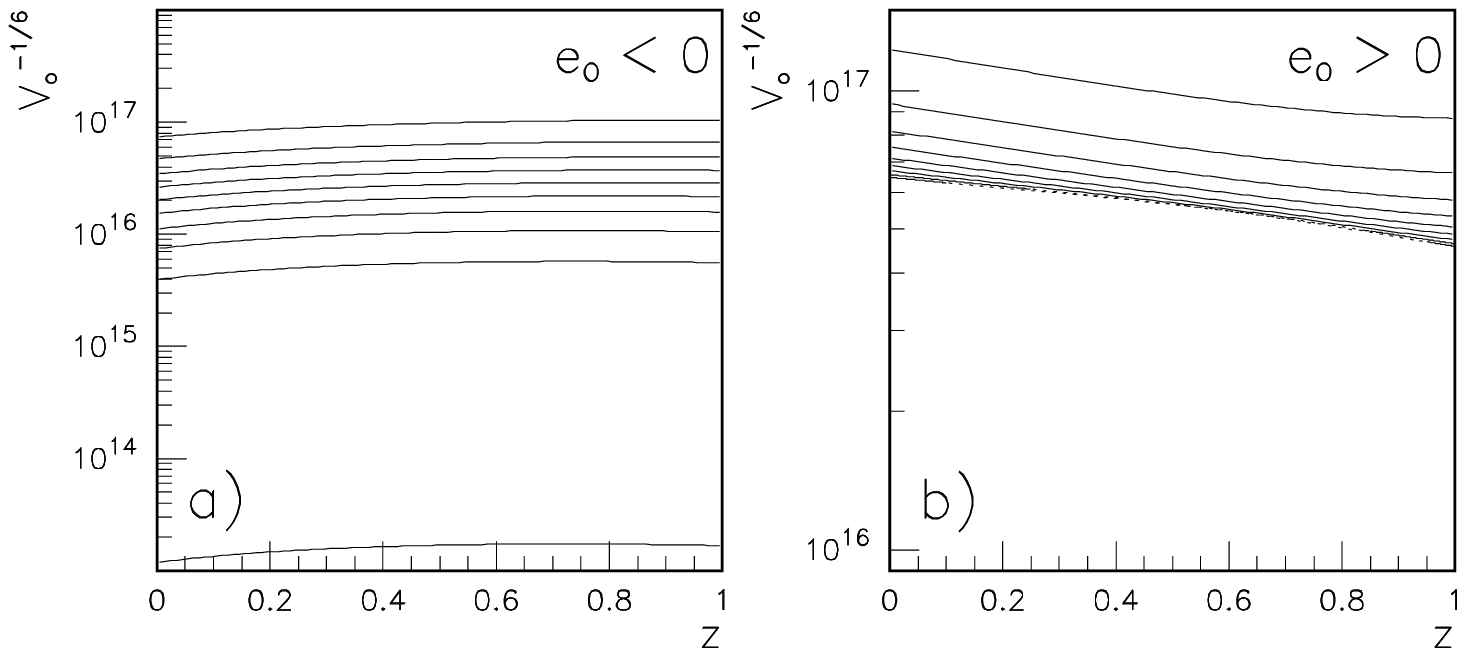}
\end{center}
\vspace*{-1cm}
\captions{Dependence of the GUT scale on the position of the five-brane along the orbifold interval for the examples discussed earlier. a) For the scenario (\ref{scenarioi}), the lines show $\sigma e_O=-0.1, -0.2, -0.3, -0.4, -0.5, -0.6, -0.7, -0.8, -0.9, -0.9999$, from top to bottom. b) For scenario (\ref{scenarioiii}), the lines show $\sigma e_O=0.1, 0.2, 0.3, 0.4, 0.5, 0.6, 0.7, 0.8, 0.9$, from top to bottom. Dotted lines for $\sigma e_O=0.8,0.9$ show forbidden regions (see Fig.\,\ref{exclusion}).
\label{scales-z}}
\end{figure}

We now analyze the scales for positive values of $\sigma e_O$, and illustrate this with the scenario described in (\ref{scenarioiii}), for which $0<\sigma e_O<5/7$. The resulting scales are represented in the right hand side of Fig.\,\ref{scalesancient}, where the vertical dotted line shows the upper bound for $\sigma e_O$. We find that now the scales are of the order of the phenomenologically favoured value for the GUT scale. For example, in this case, the lower value is reached at the largest allowed value for $\sigma e_O$, and corresponds to $V_O^{-1/6}\approx 6.3\times10^{16}$ GeV. However obtaining $3\times10^{16}$ is not as simple as in the case of negative $\sigma e_O$. We have analyzed the conditions under which this value is reached and found that for $\beta_O>0$ the resulting scale is always above this value. 
Moving the five-brane does not have a big influence on the scales in this case, although a small reduction in the value of the scale is obtained increasing $z$. This is shown in Fig.\,\ref{scales-z}, where the value of $V_O^{-1/6}$ versus the five-brane position is represented for scenarios (\ref{scenarioi}) and (\ref{scenarioiii}), for different values of $\sigma e_O$ in the allowed range.

On the contrary, when the instanton number on the observable hyperplane vanishes, i.e. $\beta_O=0$, lower scales can be obtained. The most advantageous case occurs for $\beta_O=0$, and $\beta_1=1$, but even then the predictions for the GUT scale are higher than the preferred value for a wide range in $z$. However, in this particular case, moving the five-brane towards the hidden hyperplane turns out to be crucial. It helps in obtaining the phenomenologically favoured value. 
This also offers the appealing possibility of lowering the value of the scales when the five-brane approaches even more the hidden fixed hyperplane. For example, for $z=0.999999$, intermediate scales of the order of $V_O^{-1/6}\approx10^{13}$ GeV are obtained. Of course, the fact that the value of $z$, and therefore, the VEV of the five-brane modulus, have to be so carefully tuned renders this possibility highly unnatural. Also, as discussed in the footnote \ref{foot_critic}, we do not know how to describe the theory in such a regime.

\begin{figure}[!t]
\begin{center}
\epsfig{file=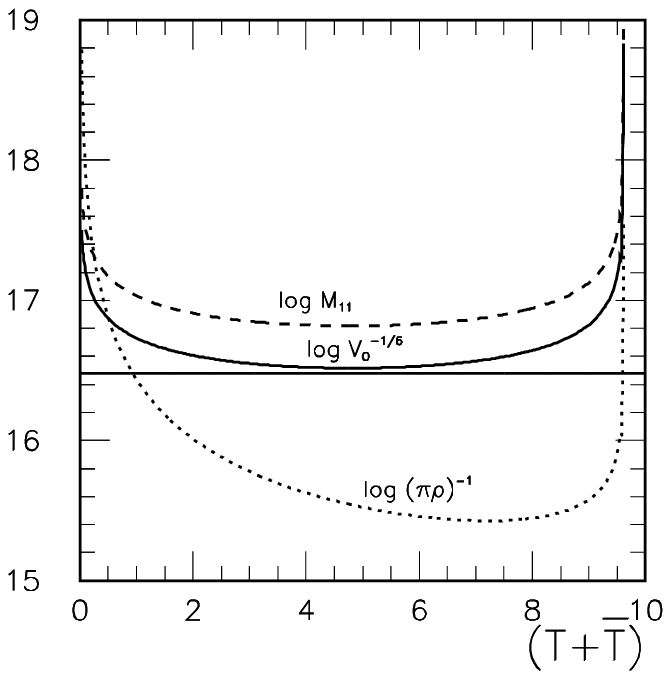}
\end{center}
\vspace*{-1cm}
\captions{Scales of the theory versus $\big(T+\bar T\big)$ for the case $b_O=0$, exemplified with scenario (\ref{scenarioii}). The phenomenologically favoured value for the GUT scale, $M_{GUT}=3\times10^{16}$ GeV, is shown as a solid line.
\label{scalesancientb0}}
\end{figure}

Let us finally study the case $b_O=0$. In this case, using (\ref{eos}), we can express (\ref{scales_volume}) in terms of $\modulit$, as well as the expressions for the rest of the scales, and we are left with:
\begin{subequations}
\label{scales-B=0}
\begin{align}
V_O^{-1/6}&=3.6\,\cdot10^{16}\left(\frac1{1+\frac{b_H}{8}\modulit}\right)^{1/2}\left(\frac 2{\rule{0pt}{2.1ex}T+\bar T}\right)^{1/2}\ {\rm GeV}\ ,
\label{volumeb0}\\
\frac{M_{11}}{V_O^{-1/6}}&=\left(2(4\pi)^{-2/3}\alpha_O\right)^{-1/6}\approx2\ ,\\
\frac{V_O^{-1/6}}{(\pi\rho)^{-1}}&=
\left(\frac1{1+\frac{b_H}{8}\modulit}\right)
\left(\frac{2.7\times10^{16}\,{\rm GeV}}{V_O^{-1/6}}\right)^2\left(8192\pi^4\alpha^3\right)^{1/2}\ .
\end{align}
\end{subequations}
Again, these are the same expressions as those found in \cite{cm99-1}. Now, it can be easily seen from (\ref{bobh}) that setting $\beta_1>0$ implies that in this case $b_H\le0$. Thus there is an upper value of the modulus $\modulit$ for which the denominator in the expression for $V_O^{-1/6}$ vanishes and the values of the scales become infinite. Larger values of $\modulit$ are therefore not valid.
An example of the scales of the theory in this case is shown in Fig.\,\ref{scalesancientb0} for the particular scenario (\ref{scenarioii}).
We see how in this case, $V_O^{-1/6}$ is of the same order as the phenomenologically favoured GUT scale. Intermediate or small scales are, nevertheless, unreachable. The upper value for $\modulit$ in this case is found for $\modulit\approx 9.8$.

We have analyzed under which conditions the value of the scale can be lowered to fit the phenomenologically preferred value. We can see from (\ref{volumeb0}) that low values of $b_H$ would let us go to higher values of $\modulit$ and this would result in lower values for the scale. Thus, the most favourable case occurs when $b_H=0$, where we have no upper bound on the value of the modulus and we would obtain in principle intermediate values for the scale. This happens for $\beta_O=0$ (and then $z=1$) or for $\beta_O=-\beta_1$ (and then $z=0$). As we see, these two are very special cases, where the five-brane coincides with either the hidden or the observable hyperplane and, as we have already discussed, the theory undergoes a small-instanton transition.
Having excluded for this reason the cases for $b_H=0$, the most advantageous cases occur for $\beta_O=-\beta_1+1$, and among them, the case $\beta_O=-1$, $\beta_1=2$ is the one which produces a lower $b_H$, which is the case already shown.

Summarizing, the phenomenologically preferred value for the GUT scale can still be obtained when there are five-branes in the bulk. In particular, this is much easier for the cases allowing for negative values of $e_O$. These cases also allow for much lower values of the scales but at the price of introducing a fine-tuning problem. For $\sigma e_O>0$ obtaining $V_O^{-1/6}=3\times10^{16}$ GeV is only possible when the instanton number of the observable hyperplane vanishes, and even then, the five-brane must be slightly close to the hidden fixed hyperplane. This case also allows to lower this scale to intermediate values if the position of the five-brane is tuned to be extremely close to $z=1$. In the case where $b_O=0$ the value $V_O^{-1/6}=3\times10^{16}$ GeV is not reached. The lower values of the scales corresponding to the case $\beta_O=-1$ and $\beta_1=2$ are, however, very close to this value.


\section{Soft terms}
\setcounter{equation}{0}
\label{sec_softterms}

The soft supersymmetry-breaking terms are evaluated by applying the standard tree-level formulae \cite{sw83} (see \cite{bim97} for a review) to the expressions (\ref{kahlerfunction}) and (\ref{gaugekinetic}), and using the superpotential $W=d_{pqr}C_O^pC_O^qC_O^r$. Due to the compactification on a Calabi-Yau with only one modulus $T$ that we are using, the soft terms are universal and read:
\begin{subequations}
\label{softterms}
\begin{align}
\begin{split}\label{escalares}
m^2&=(m_{3/2}^2+V_0)-\frac{1}{(3+e_O)^2}\left\{
|F^S|^2{\frac {e_O\left (6+e_O\right )}{{\dilaton}^{2}}}
\right.\\
&+2\,{\rm Re}({F^S\bar F^{\bar T}}) \frac{-3\,e_O\left(2\,z\,\frac{\beta_1}{b_O}\left (1-z\right )+1\right)}
{{\dilaton}{\modulit}}\\
&+|F^T|^2 \frac{9+6e_O-4e_O^2z\frac{\beta_1}{b_O}+6z^2\frac{\beta_1}{b_O}e_O(1+e_O)-4e_O^2\frac{{\beta_1}^2}{b_O^2}z^2(1-z)^2}{{\modulit}^{2}}\\
&+2\,{\rm Re}({F^S\bar F^{\bar Z}}) \frac{6\left(1-z\right)}{{\dilaton}^{2}}\\
&+2\,{\rm Re}({F^T\bar F^{\bar Z}})\frac{2\,e_O-2\,z\,\left (3+2\,e_O\right )+4\,e_O\,z\frac{\beta_1}{b_O}\,\left(1-z\right)^2}{{\dilaton}{\modulit}}\\
&\left.+|F^Z|^2 \left(2\,{\frac {3+e_O}{\dilaton\modulit\beta_1}}-\frac{4}{\dilaton^2}\,\left (1-z\right)^{2}\right)\right\}\ ,
\end{split}\\
\begin{split}\label{trilinear}
A&=-\frac1{\left(3+e_O\right)}\frac1{\left(1-{\frac {{z}^{2}\beta_1\,e_O}{b_O}}\right)}
\left\{\frac{F^S}{\dilaton}\left(3-2\,e_O+3\,{\frac {{e_O}^{2}{z}^{2}\beta_1}{b_O}}\right)\right.\\
&+\frac{F^T}{\modulit}\left(3\,e_O-2\,{\frac {{e_O}^{2}{z}^{2}\beta_1}{b_O}}+3\,{\frac {e_Oz\beta_1\,\left (2-z\right )}{b_O}}-6\,{\frac {{e_O}^{2}{z}^{3}{\beta_1}^{2}\left (1-z\right )}{{b_O}^{2}}}\right)\\
&-\left.\frac{F^Z}{\dilaton}\left(6+2\,e_Oz-6\,{\frac {e_O\beta_1\,{z}^{2}\left (1-z\right )}{b_O}}\right)\right\}\ ,
\end{split}\\
\begin{split}\label{gaugino}
M&=\frac{F^S+F^T\left(\beta_O+\beta_1\right)-2F^Z}{\dilaton+\modulit\left(\beta_O+\beta_1\left(1-2z\right)\right)}\ ,
\end{split}
\end{align}
\end{subequations}
where $m$, $A$, and $M$ stand for scalar masses, trilinear parameters and gaugino masses, respectively. 
We will use a parameterization, introduced in \cite{bim93-1}, for the auxiliary $F$-terms in order to know which field ($S$, $T$, or $Z$) plays a predominant role in the process of supersymmetry breaking.
These $F$-terms appear in the expression for the tree level scalar potential, $V_0$, and are parameterized in such a way that:
\begin{equation}
V_0=\bar F^{\bar n} K_{\bar nm}F^m-3m_{3/2}^2\ .
\end{equation}
where $K_{m\bar n}=\frac{\partial^2  K}{\partial Y^m\bar\partial \bar Y^{\bar n}}$ is the \kah\ metric, and $Y^m$ stands for the chiral fields $S$, $T$, $Z$.
The above condition is fulfilled by setting:
\begin{equation}
F^m=\sqrt{3}\,C\,m_{3/2}\,P^{m\bar n}\Theta_{\bar n}\ ,
\label{parameterization}
\end{equation}
where $C^2=1+\frac{V_0}{3m^2_{3/2}}$, $P^{m\bar n}$ is a matrix which satisfies
\begin{equation}
P^\dagger KP=\mathbb{I}\ ,
\end{equation}
and $\Theta_n$ are complex numbers which satisfy the constraint
\begin{equation}
\sum_n\Theta^*_n\Theta_n=1\ . 
\end{equation}
This last constraint allows us to write
\begin{eqnarray}
\Theta_S&=&\sin\theta\cos\Theta\ e^{-i\gamma_S}\ ,\nn\\
\Theta_T&=&\cos\theta\cos\Theta\ e^{-i\gamma_T}\ ,\nn\\
\Theta_Z&=&\sin\Theta\ e^{-i\gamma_Z}\ ,
\label{goldstinos}
\end{eqnarray}
where we have introduced two goldstino angles, $\theta$ and $\Theta$.

Taking into account the current experimental limits, we will assume $V_0=0$ (i.e. $C=1$) and fix the phases $\gamma_{S,T,Z}=0$. This implies that $\theta$ and $\Theta$ must vary in a range $[0,2\pi)$ and $[0,\pi)$, respectively. As a result of parameterization (\ref{parameterization}) for the $F$-terms, there exist the following symmetries in the expressions for the soft terms. Under the shift $\Theta\rightarrow\Theta+\pi$ the soft terms transform as $m\rightarrow m$, $M\rightarrow-M$, and $A\rightarrow-A$. Also, under the shifts $\Theta\rightarrow \pi-\Theta$, and $\theta\rightarrow \theta+\pi$, $m$, $M$, and $A$ remain invariant. We can therefore analyze the region $\Theta\in[0,\pi/2]$ and the rest of the figures can be easily deduced.

\begin{figure}[!t]
\begin{center}
\epsfig{file=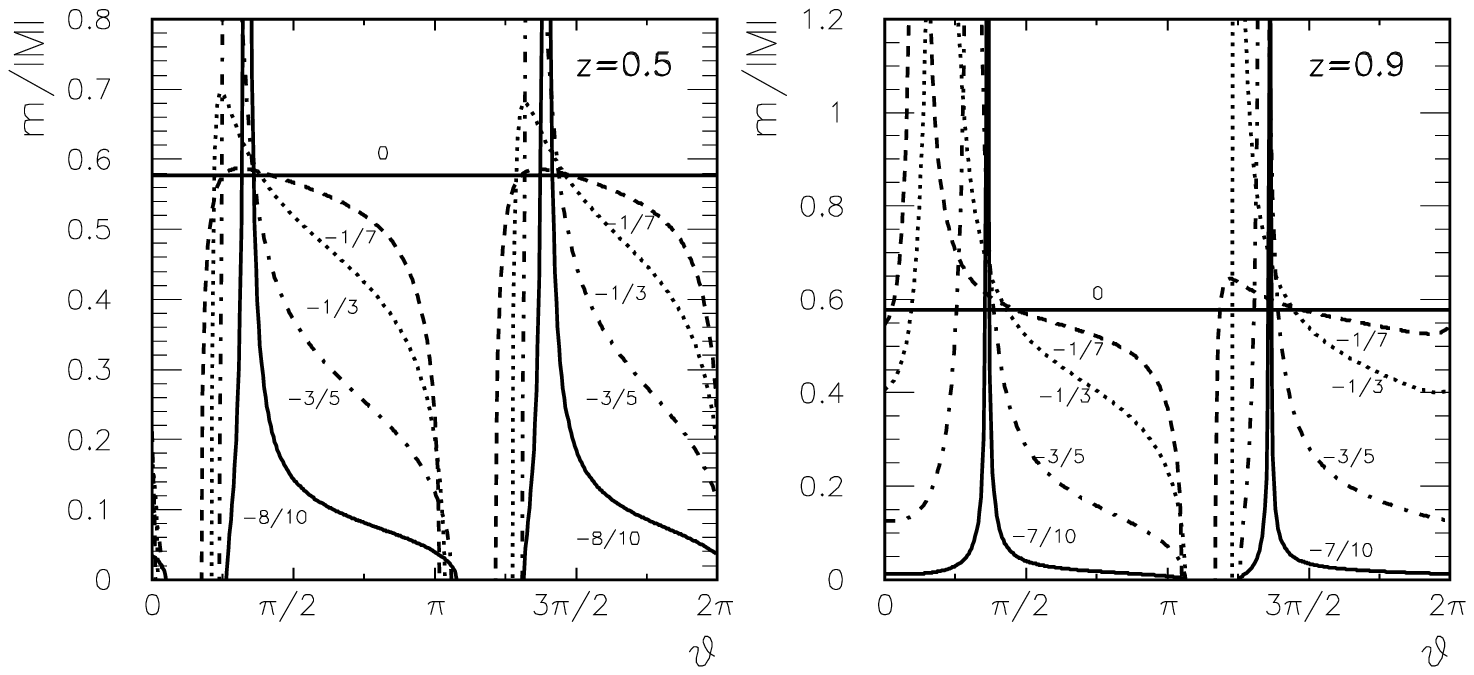}
\end{center}
\vspace*{-1cm}
\captions{Ratio $m/|M|$ versus $\theta$ for different values of $e_O$ and $\Theta=\pi/4$ for a generalization of scenario (\ref{scenarioi}) for two different five-brane positions, $z=0.5,\ 0.9$.
\label{softratiose<0}}
\end{figure}

Let us now consider some particular examples. We will first concentrate on negative values of $e_O$ and illustrate this with the scenario described in (\ref{scenarioi}).
\begin{figure}[!t]
\begin{center}
\epsfig{file=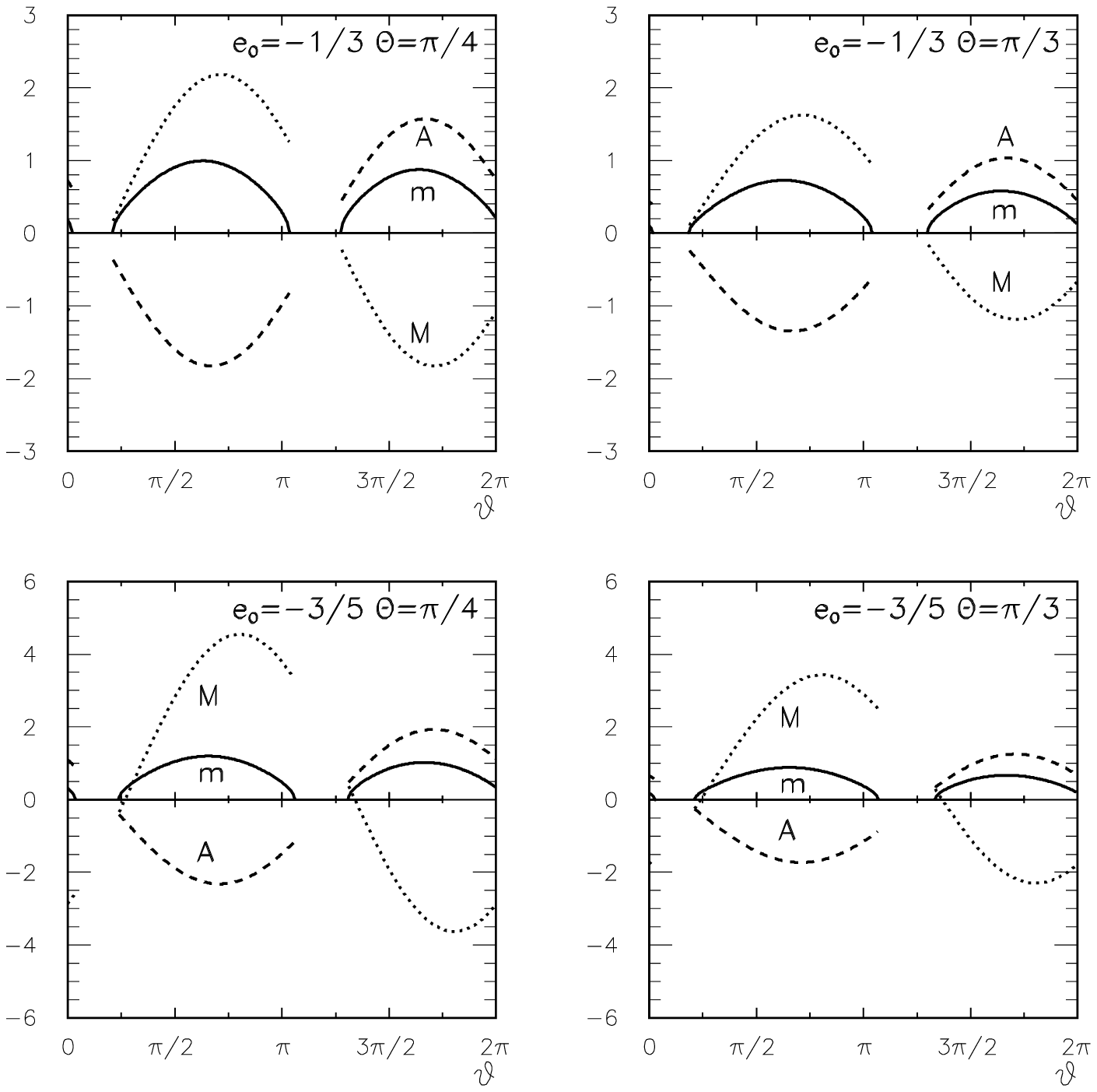}
\end{center}
\vspace*{-1cm}
\captions{Soft parameters in units of $m_{3/2}$ versus $\theta$ for different values of $e_O$ and $\Theta$ in the allowed range for scenario (\ref{scenarioi}).
\label{softtermse<0}}
\end{figure}
The soft terms for different values of $e_O$ and $\Theta$ are shown in Fig.\,\ref{softtermse<0}. We can see that there is a region in $\theta$ which has been excluded in order to avoid negative values of $m^2$. This region is bigger for smaller (more negative) values of $e_O$. We also find how the amplitude in $M$ increases when $e_O$ is more negative. This is due to the following fact. Expression (\ref{gaugino}), when (\ref{sigmadef}) is used, can be shown to be $M\sim(1+\sigma e_O)^{-1}$. Therefore, $M$ grows as $\sigma e_O$ approaches $-1$, which corresponds to $e_O$ approaching its lower limit ($-\frac78$ in this case). One more feature that can be seen in Fig.\,\ref{softtermse<0} is that when $e_O$ is more negative, there are small regions in $\theta$ where scalar masses larger than gaugino masses can be obtained. This appears, for example in the figures corresponding to $e_O=-\frac35$. This is more clearly shown in Fig.\,\ref{softratiose<0}, where the ratio $m/|M|$ is represented for several values of $e_O$ and $\Theta=\pi/4$  inside the allowed range determined in (\ref{scenarioi}) for $z=0.5$ and also a generalization of this scenario for $z=0.9$. The case $e_O\rightarrow0$ corresponds to the weakly coupled heterotic string limit, as we will explain below, and therefore satisfies the sum rule $3m^2=M^2$ and is represented with a straight line. For the rest of the cases, this is in general not true.

For low values of $z$ all of this is very similar to what one obtains without five-branes, as it can be seen in Figs.\,4 and 5 of \cite{cm99-1}. However, all the aforementioned features are more easily obtained when five-branes are introduced. Compare, for example, the case $e_O=-\frac35$ in Fig.\,\ref{softtermse<0} here and Fig.\,4 in \cite{cm99-1}. We find that $m>|M|$ is possible now for a wider range in $\theta$. Furthermore, when the five-brane is near the hidden hyperplane the effects of the five-brane are more important and, in particular, scalar masses larger than gaugino masses are much more common. This is shown in Fig.\,\ref{softratiose<0} for $z=0.9$, where one can see that for this to happen we do not need $e_O$ to be close to its lower limit.

Let us concentrate now on the case $e_O\rightarrow 0$. As we have already discussed, the behavior of the scales in this limit (see Fig.\,\ref{scalesancient}) indicates that that we should recover the weakly coupled limit. Let us analyze this from the point of view of the soft terms. It can be seen from (\ref{kahlerfunction}) that the \kah\ metric is almost diagonal and the expressions of the $F$-terms can be written as:
\begin{eqnarray}
F^S&\sim&4\,\sqrt{3}\,\Theta_S\,m_{3/2}\ ,\nn\\
F^T&\sim&\modulit\,\Theta_T\,m_{3/2}\ ,\nn\\
F^Z&\sim&\sqrt{6\beta^1\modulit}\,\Theta_Z\,m_{3/2}\ .
\end{eqnarray}
Inserting these in the expressions of the soft terms and taking the limit $e_O\rightarrow 0$ we are left with the following expressions:
\begin{eqnarray}
-A=M&=&\sqrt{3}\,\Theta_S\,m_{3/2}\ ,\nn\\
m&=&\Theta_S\,m_{3/2}\ .
\end{eqnarray}
This is what one obtains in the weakly coupled heterotic string once the parameter $\Theta_S$ is absorbed in the definition of the gravitino mass. Therefore in this limit we recover the sum rule $3m^2=M^2$.

\begin{figure}[!t]
\begin{center}
\epsfig{file=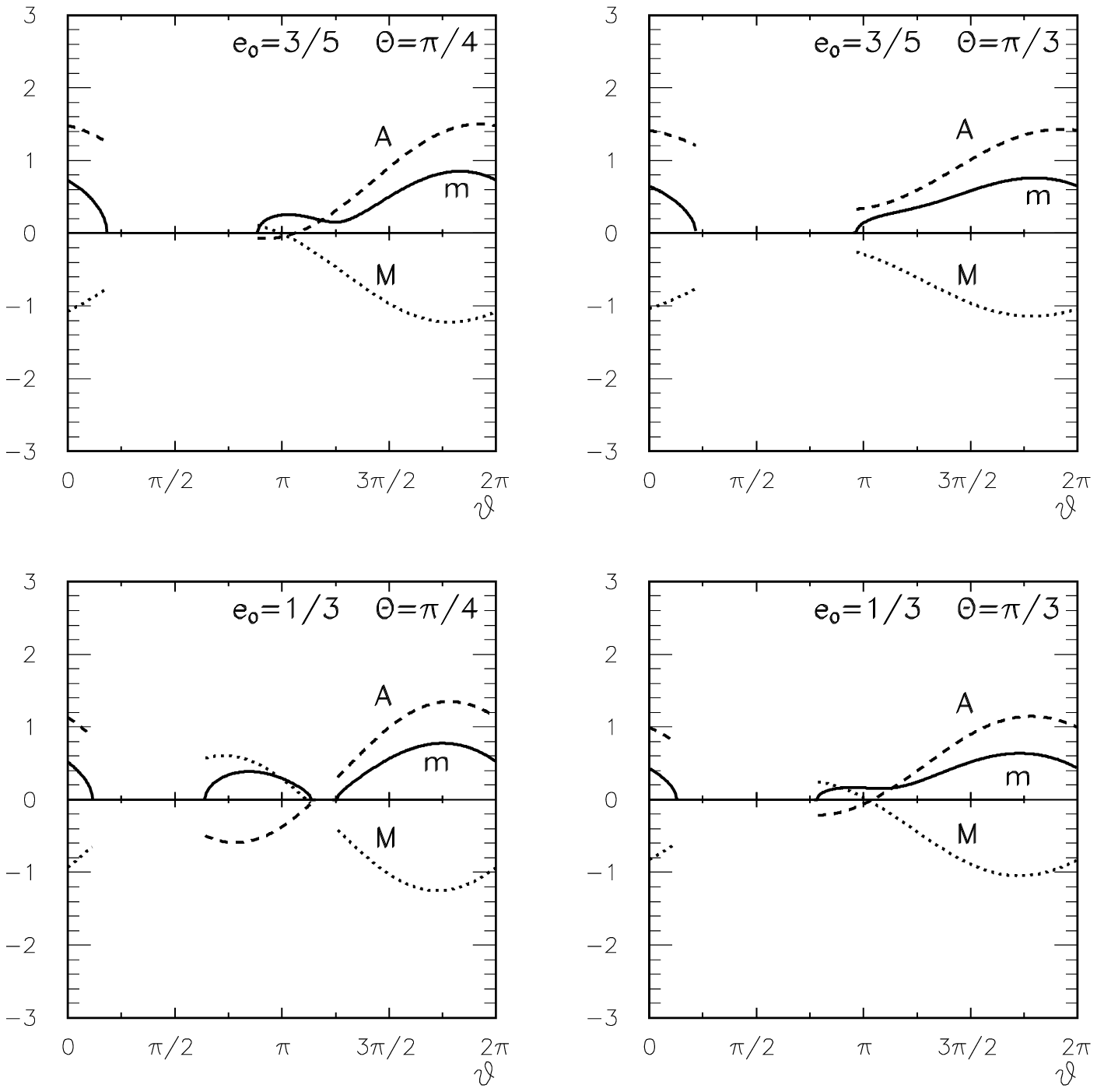}
\end{center}
\vspace*{-1cm}
\captions{Soft parameters in units of $m_{3/2}$ versus  $\theta$ for different values of $e_O$ and $\Theta$ in the allowed range for scenario (\ref{scenarioiii}).
\label{softtermse>0}}
\end{figure}

\begin{figure}[!t]
\begin{center}
\epsfig{file=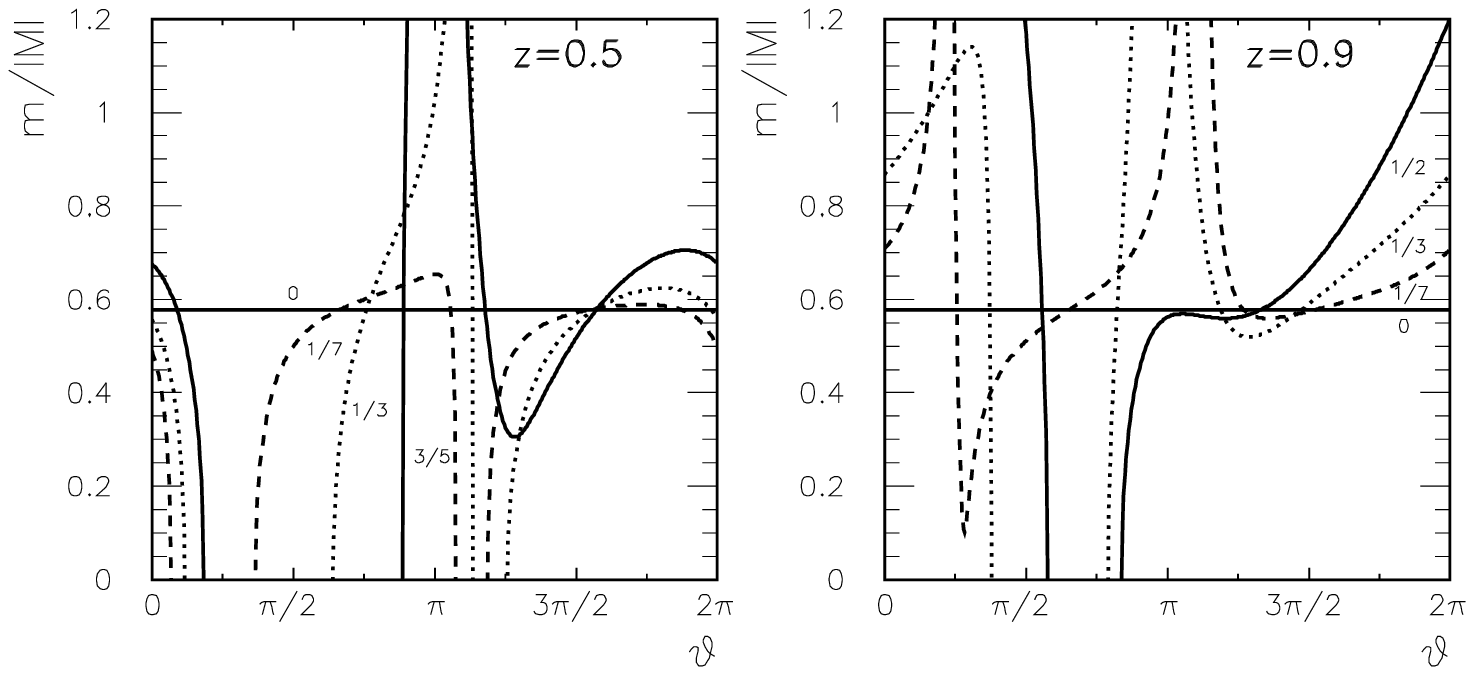}
\end{center}
\vspace*{-1cm}
\captions{Ratio $m/|M|$ versus $\theta$ for different values of $e_O$  and $\Theta=\pi/4$ for a generalization of scenario (\ref{scenarioiii}) for two different five-brane positions, $z=0.5,\ 0.9$.
\label{softratiose>0}}
\end{figure}

Let us now consider positive values of $e_O$.
We will work with the scenario defined in (\ref{scenarioiii}).
The soft terms are represented in Fig.\,\ref{softtermse>0} for some values of $\Theta$ and $e_O$ inside the allowed region. We find again forbidden regions in $\theta$ for having negative $m^2$ and notice that also in this case there are some ranges in $\theta$ where scalar masses are larger than gaugino masses. The ratio $m/|M|$ for several values of $e_O$ and $\Theta=\pi/4$ is shown in  Fig.\,\ref{softratiose>0} for two positions of the five-brane ($z=0.5,\ 0.9$), where one can see that obtaining $m>|M|$ is much easier than in the case with negative $e_O$. The ranges in $\theta$ are now much larger. These ranges have a strong dependence on $e_O$ and $\Theta$ as is also clear in this figure. We have also checked that the closer the five-brane to the hidden hyperplane is the wider these regions become. The weakly coupled limit $e_O\rightarrow 0$ is also shown and again it is the only case for which the sum rule $3m^2=M^2$ holds.

The standard embedding case without five-branes, where $e_O>0$, was studied in \cite{ckm97-1} (the non-standard embedding case for $e_O>0$ has also the same results \cite{cm99-1}). The soft terms in the cases with and without five-branes clearly differ, as can be seen when comparing Fig.\,\ref{softtermse>0} here to Fig.\,1 of \cite{ckm97-1}. Not only are the allowed regions having $m^2>0$ different, but, what is more important, those regions with $m>|M|$ could not be obtained without five-branes.
The case $e_O>0$ is therefore much more sensitive to the presence of five-branes than the case with negative $e_O$.

\begin{figure}
\begin{center}
\epsfig{file=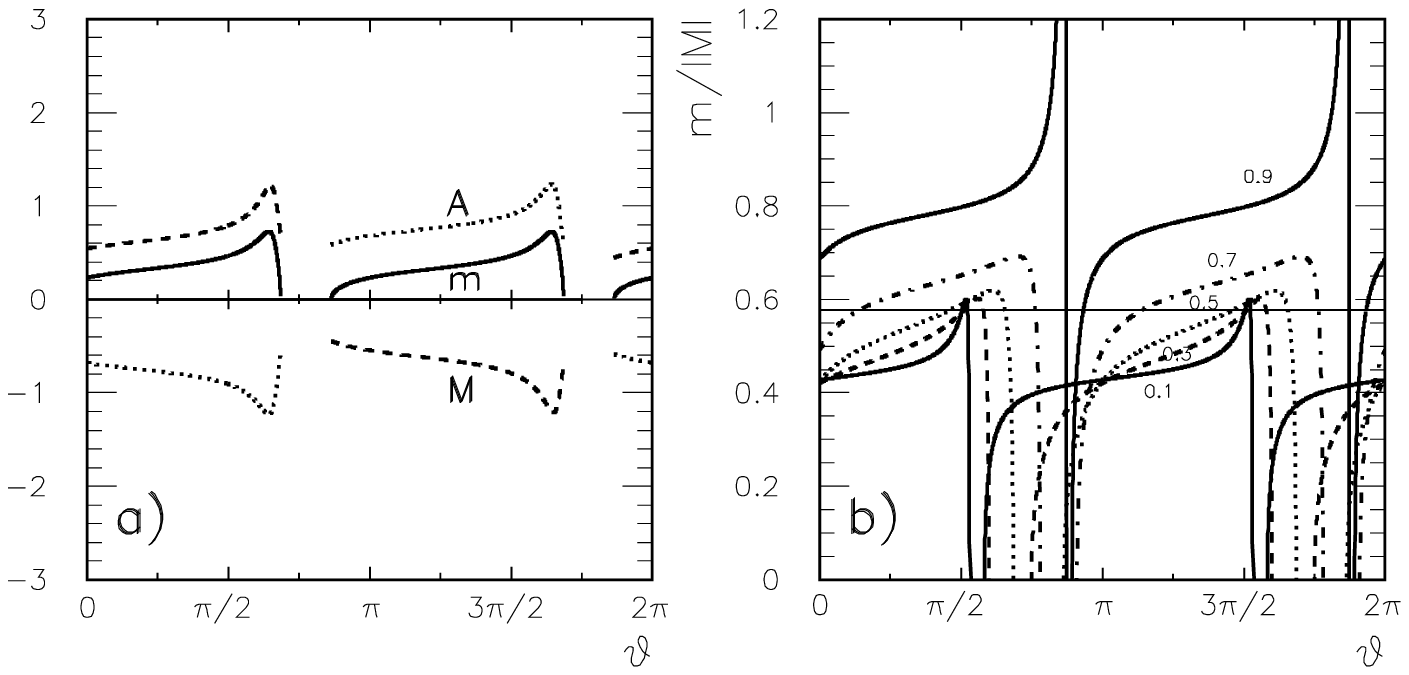}
\end{center}
\vspace*{-1cm}
\captions{a) Soft parameters in units of $m_{3/2}$  versus $\theta$ for scenario (\ref{scenarioiii}) for the case $e_O=0.5$ when $F^Z=0$. b) The ratio $m/|M|$ is shown for a generalization of scenario (\ref{scenarioiii}), keeping $e_O=0.5$ for several values of $z$.
\label{braneless}}
\end{figure}

It is interesting to analyze the case where the five-brane does not contribute to the breaking of supersymmetry, i.e., the $F$-term associated to the five-brane vanishes, $F^Z=0$. This condition, once the goldstino angles (\ref{goldstinos}) are introduced, leaves us with a single parameter to play with, namely one of the goldstino angles. As we can see from expressions (\ref{softterms}), the presence of the five-brane affects the soft terms, even after considering this special case. This will be illustrated for a particular example in scenario (\ref{scenarioi}), with $e_O=0.5$.
For this case, the resulting $F$-terms, satisfying $F^Z=0$ can be straightforwardly computed and are (modulo transformations which can be absorbed as translations of the goldstino angles):
\begin{eqnarray}
F^S&\approx&\sqrt{3}\left(2.84\ \sin\theta\cos\Theta-0.24\ \cos\theta\cos\Theta-0.018\ \sin\Theta\right)m_{3/2}\ ,\nn\\
F^T&\approx&\sqrt{3}\left(0.056\ \sin\theta\cos\Theta+0.25\ \cos\theta\cos\Theta-0.61\ \sin\Theta\right)m_{3/2}\ ,
\end{eqnarray}
with
\begin{equation}
\Theta\approx\arctan\left(-\frac{0.63\ \sin\theta+1.08\ \cos\theta}{0.13}\right)\ .
\end{equation}
We plot this case for the whole range in $\theta$ in Fig.\,\ref{braneless}a and notice how the structure of the soft terms is altered by the introduction of the five-brane despite the fact that this does not contribute to the breaking of supersymmetry. As in previous examples, the effect of the five-brane is much more important when its position is near the hidden hyperplane. This is illustrated in Fig.\,\ref{braneless}b, where the ratio $m/|M|$ versus $\theta$ is shown for the scenario (\ref{scenarioiii}), generalized for different positions of the five-brane. In all cases the value $e_O=0.5$ has been taken (as Fig.\,\ref{exclusion} shows, this value is inside the allowed region for all the values of $z$). This figure shows that scalar masses larger than gaugino masses are also possible in this limit but only if the five-brane is close to the hidden hyperplane. In our case this happens for $z\gsim0.85$.
For negative values of $e_O$ the above discussion still holds. Although now scalar masses larger than gaugino masses can be obtained for all values of $z$, the range in $\theta$ where this happens increases for larger $z$.
We have also analyzed the limit $F^Z=0$ for negative values of the parameter $e_O$, using scenario (\ref{scenarioi}) as an example, finding the same qualitative results.

\begin{figure}[!t]
\begin{center}
\epsfig{file=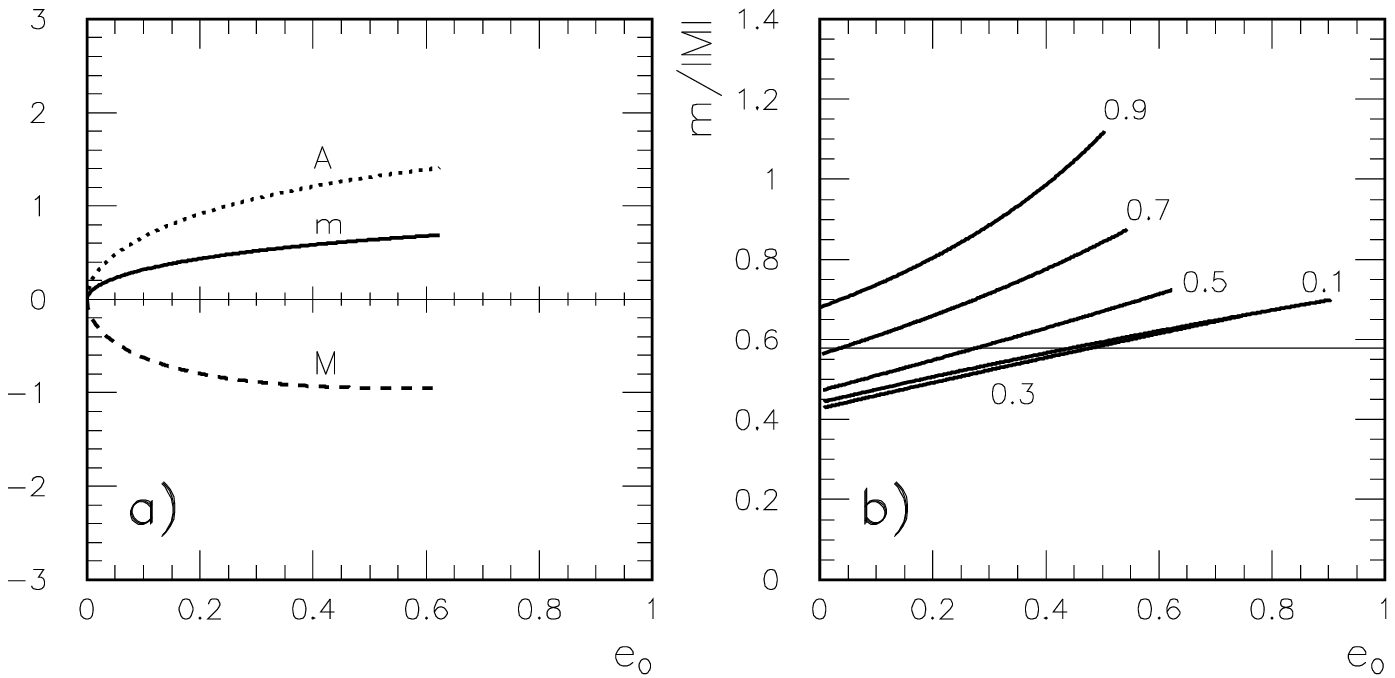}
\end{center}
\vspace*{-1cm}
\captions{a) Soft terms in units of the gravitino mass versus $e_O$ for the five-brane dominated supersymmetry-breaking in scenario (\ref{scenarioiii}). b) Ratio $m/|M|$ in the five-brane dominated case for a generalization of (\ref{scenarioiii}) for different values of the five-brane position over the allowed ranges of $e_O$.
\label{fsft0_iii}}
\end{figure}

Finally, let us analyze the case where the five-brane modulus is the only one responsible for the breaking of supersymmetry. This happens when both $F^S$ and $F^T$ vanish.
Imposing these conditions leaves us a single point (modulo symmetry transformations) in the $(\theta,\Theta)$ parameter space.
Again, we will illustrate this in scenario (\ref{scenarioi}). The conditions $F^S=0$ and $F^T=0$ are fulfilled in this case for the following choices of goldstino angles:
\begin{eqnarray}
\theta\approx0.088\quad&\rightarrow&\quad \Theta\approx0.40\quad(+\pi)\ ,\nn\\
\theta\approx0.088+\pi\quad&\rightarrow&\quad \Theta\approx-0.40\quad(+\pi)\ ,
\label{fsft0thetas}
\end{eqnarray}
and $F^Z$ is determined in terms of these as:
\begin{equation}
F^Z\approx\sqrt{3}\left(0.63\ \sin\theta\cos\Theta+1.08\ \cos\theta\cos\Theta+0.13\ \sin\Theta\right)m_{3/2}\ .
\end{equation}
Thus we are left with only four points, which are in fact related by the symmetry transformations we have already discussed.  
For the choice $\theta\approx0.088,\Theta\approx0.40$ 
the corresponding soft terms are:
\begin{eqnarray}
m^2&\approx&0.41\ m_{3/2}^2\ ,\nn\\ 
M&\approx&-0.95\ m_{3/2}\ ,\nn\\ 
A&\approx&1.31\ m_{3/2}\ . 
\label{fsft0soft}
\end{eqnarray}
For the other choices we can use the relations under transformations of the goldstino angles we have already discussed.
For this explicit example, we have also studied the effect of the variation of $e_O$ on the soft terms. This is illustrated in Fig\,\ref{fsft0_iii}a where the soft terms are shown for the allowed range in $e_O$. There is only one more parameter to play with in this case, namely, the five-brane position. We have analyzed whether scalar masses larger than gaugino masses could be obtained in this five-brane dominated case. This is shown for scenario (\ref{scenarioiii}) in Fig.\,\ref{fsft0_iii}b, where the whole range of five-brane positions has been analyzed and for each $z$, $e_O$ is varied in all its allowed range (see Fig.\,\ref{exclusion}b). We find that in general, scalar masses are lower than gaugino masses, but the ratio $m/|M|$ increases for larger values of $z$. Eventually scalar masses larger than gaugino masses can be obtained for large values of $z$ (in this case $z\gsim 0.85$) and $e_O$ near its upper limit. 

\begin{figure}[!t]
\begin{center}
\epsfig{file=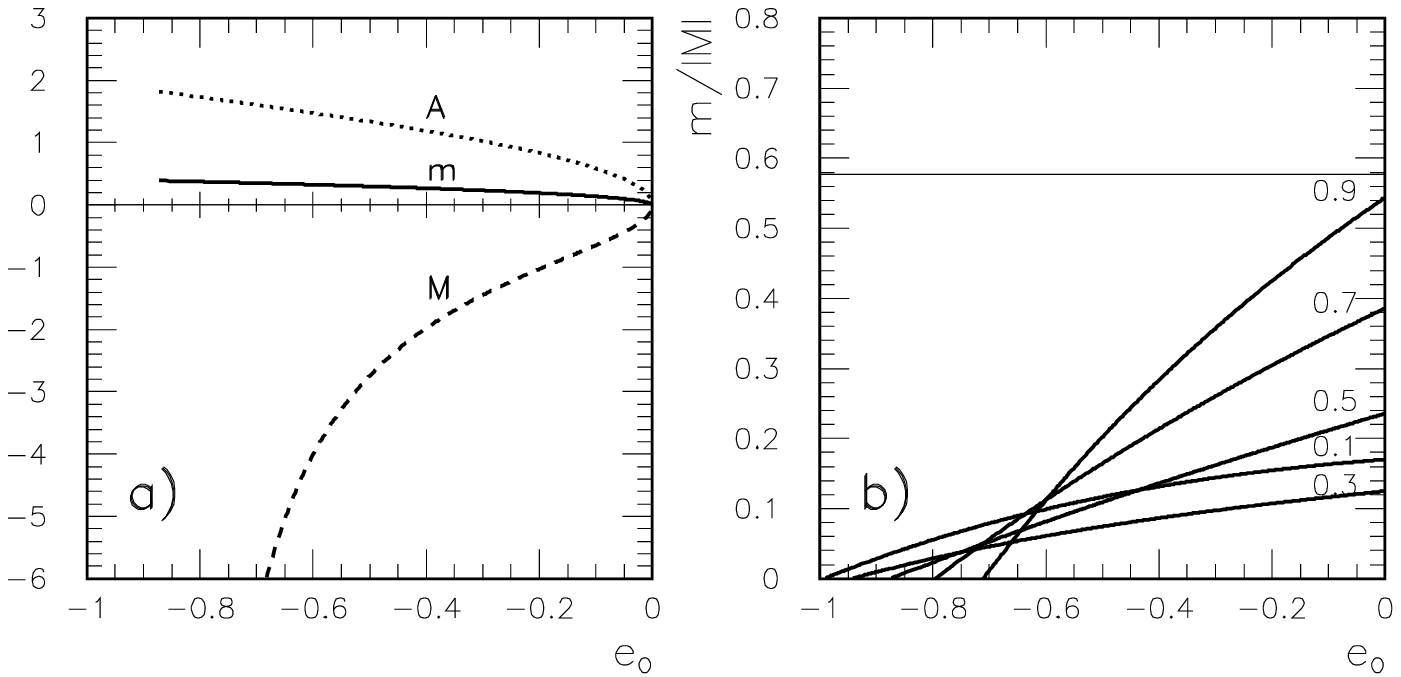}
\end{center}
\vspace*{-1cm}
\captions{a) Soft terms in units of the gravitino mass versus $e_O$ for the five-brane dominated supersymmetry-breaking in scenario (\ref{scenarioi}). b) Ratio $m/|M|$ in the five-brane dominated case for a generalization of (\ref{scenarioi}) for different values of the five-brane position over the allowed ranges of $e_O$.
\label{fsft0_i}}
\end{figure}

We can also analyze the five-brane dominated limit for the cases with negative values of the $e_O$ parameter, as for instance, in the scenario (\ref{scenarioi}). The soft terms in this case are shown in Fig.\,\ref{fsft0_i}a for the allowed range of values for $e_O$. It can be seen in the figure how $M$ diverges when $e_O$ approaches its lower limit. This behavior can be easily understood by analyzing expression (\ref{gaugino}). We see there how $M$ is proportional to $(1-\sigma e_O)^{-1}$, and as the lower limit for $e_O$ in this case corresponds to $\sigma e_O\rightarrow-1$, $M$ diverges. It is clear from this that finding regions with scalar masses larger than gravitino masses is more difficult in this case. The ratio between these two masses is plotted in Fig.\,\ref{fsft0_i}b for several positions of the five-brane along the orbifold interval. We find again that this ratio tends to increase when the five-brane approaches the hidden hyperplane. However, now, contrary to what we found for positive $e_O$, scalar masses larger than gaugino masses cannot be obtained.

Summarizing, the presence of five-branes alter the structure of the soft terms. First, they introduce new degrees of freedom parameterizing its charge and position and also a new goldstino angle. Thus, their analysis becomes much more involved. Second, new qualitative features can be found. In particular, scalar masses larger than gaugino masses are possible and much more common than in the cases without five-branes. After a thorough analysis of all the parameter space we find that this feature is more easily obtained when the five-brane is close to the hidden hyperplane, both for positive and negative values of $e_O$. Several particular cases have also been analyzed with similar results. For instance, the case when the five-brane does not contribute to the breaking of supersymmetry still shows an important dependence on the five-brane parameters and $m/|M|>1$ is obtained for large $z$. The five-brane dominated scenario has also been studied. In this case, only for positive $e_O$ are scalar masses larger than gaugino masses found, although in all the cases, the ratio $m/|M|$ increases for large $z$.

\section{Muon anomalous magnetic moment}
\setcounter{equation}{0}
\label{sec_muon}

In the previous sections we have performed a complete analysis of the structure of the scales and soft supersymmetry-breaking terms of heterotic M-theory with five-branes. These results allow us to make some predictions on low energy observables.
In particular, the supersymmetric contribution to the muon anomalous magnetic moment, \asusy, can be evaluated and the theoretical results compared with those of a recent measurement in the E821 experiment at the BNL \cite{muong201-1}. It is worth recalling that this measurement implied apparently a $2.6\sigma$ deviation from the SM predictions. In particular, taking a $2\sigma$ range around the E821 central value, one would have $11\times10^{-10}\le\amu(E821)-\amu(SM)\le75\times10^{-10}$. However, recent theoretical computations \cite{kn01-1,knpr01-1,hk01-1,bcm01-1,bpp01-1} have shown that a significant part of this discrepancy was due to the evaluation of the hadronic light-by-light contribution \cite{bpp95-1,bpp95-2,hks95-1,hks96-1,hk97-1}. As a consequence, the new constraint on any possible supersymmetric contribution is $-7\times10^{-10}\le\amu(E821)-\amu(SM)\le57\times10^{-10}$ at $2\sigma$ level.

In this section, the theoretical predictions for \asusy\ will be evaluated first in the standard and non-standard embedding cases.  We will later on focus our attention on the vacua with five-branes. 
We will assume that the MSSM can be obtained by the breaking of the $E_8$ gauge group of the observable hyperplane. It will be considered that the resulting matter content is the same as in the MSSM, and therefore, the unification scale should also be around $M_{GUT}\approx3\times10^{16}$ GeV. Thus we will ignore the possibility of lowering the GUT scale.

We will impose the experimental lower limits on the masses of the supersymmetric particles coming from LEP searches \cite{wwwsusywg_charginos,wwwsusywg_sleptons}. Another important constraint comes from the measurement of the branching ratio of the rare decay \bsg\ at CLEO \cite{cleo_bsg01} and BELLE \cite{tajima01-1} ,
$2\times 10^{-4}\le BR(b\rightarrow s\gamma)\le4.1\times 10^{-4}$. 
Finally, we will impose that the lightest neutralino, $\tilde\chi_1^0$, is the lightest supersymmetric particle, so that it constitutes a dark matter candidate (we will come back to this point in the next section). 
Cosmological constraints such as the observational bounds on the relic density $0.1\lesssim\Omega_{\tilde\chi^0_1}h^2\lesssim 0.3$ will not be applied, being these dependent on assumptions about the evolution of the early Universe (see \cite{kmt02-1} and references therein).

Let us first concentrate on standard and non-standard embeddings without five-branes.
Before entering in details, let us discuss the parameter space of these scenarios. On the one hand, as usual in supersymmetric theories, the requirement of correct electroweak breaking leaves us (modulo the sign of $\mu$) with the following parameters: the soft breaking terms (scalar and gaugino masses, and trilinear parameters), and $\tan\beta$. The soft terms are expressed in terms of three free parameters: the gravitino mass $m_{3/2}$, the Goldstino angle, $\theta$, and the parameter $\epsilon_O$. This last parameter is equivalent to $e_O$ in the limit without five-branes, and can therefore be obtained from (\ref{eoeh}) and (\ref{bobh}) by taking $\beta_1=0$. 

\begin{figure}[!t]
\begin{center}
\epsfig{file=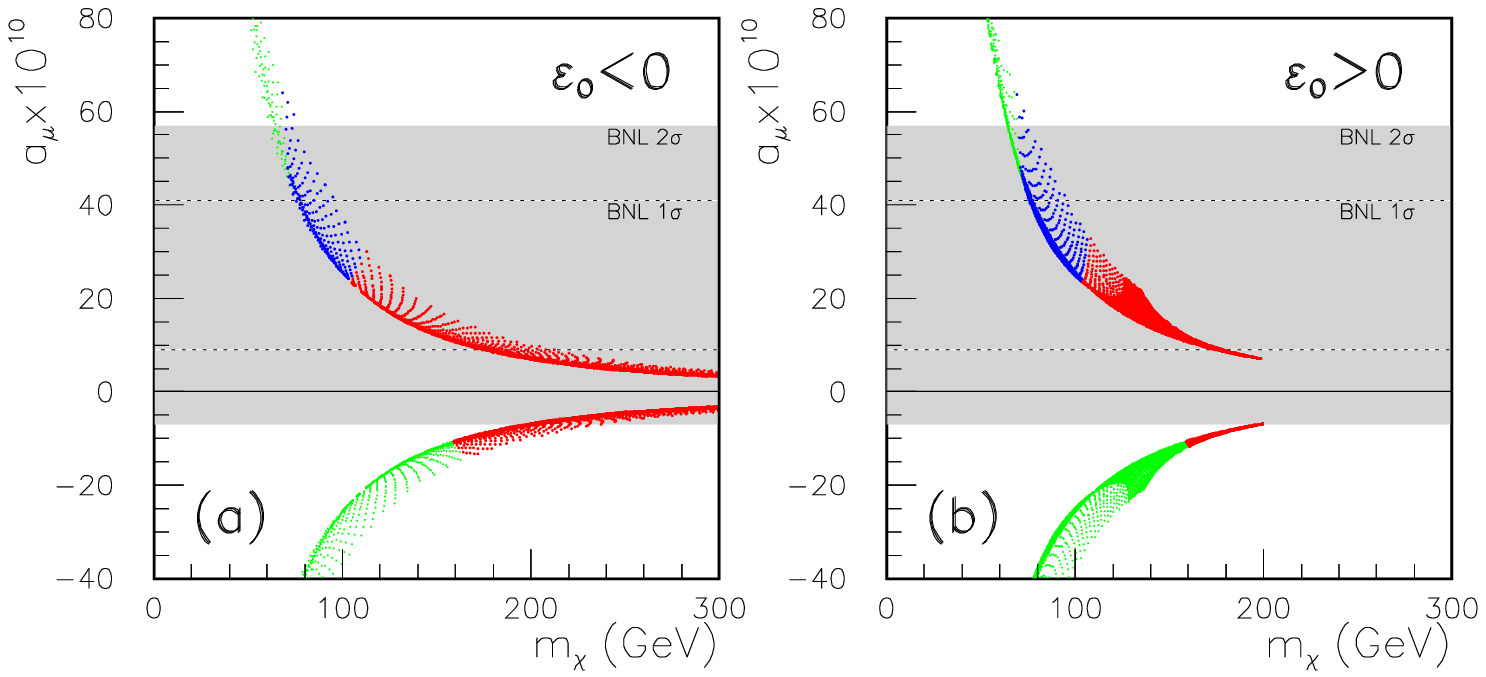}
\end{center}
\vspace*{-0.5cm}
\captions{a) Supersymmetric contribution to $\amu$ versus the neutralino mass in non-standard embedding without five-branes for $-0.6\le\epsilon_O\le-0.1$. b) The same for the standard embedding with $0.1\le\epsilon_O<1$. In both cases only the big (red and blue) dots fulfill both \bsg\ and \asusy\ constraints. The red ones correspond to points with $m_h\ge114$ GeV, whereas the blue ones correspond to points with $91\le m_h\le114$ GeV.\label{nofbg2.eps}}
\end{figure}

In the standard embedding case $M\sim -A$, as can be seen in Fig.\,1 of \cite{ckm97-1}. The non-standard embedding only deviates from this behavior when $\epsilon_O\to -1$, but in this case the phenomenologically favoured value of the GUT scale is not recovered. If we demand that this scale is obtained we have to consider moderate values of $\epsilon_O$ and therefore we also have $M\sim-A$ in most of the cases (see Fig.\,4 of \cite{cm99-1}). In fact, this constrains the values of $\epsilon_O$ that will be allowed. In particular, from Fig.\,2 of \cite{cm99-1}, we find that a sensible choice is $-0.6\le\epsilon_O\le-0.1$ for non-standard embeddings and $0.1\le\epsilon_O<1$ for the standard one. These two cases are depicted in Fig.\,\ref{nofbg2.eps} here, where \asusy\ is plotted versus the lightest neutralino mass. 
The whole allowed range for the goldstino angle, $\theta$, has been explored (with $\beta_O=-1$ and $\beta_O=1$ for non-standard and standard embeddings, respectively). The gravitino mass is set to $m_{3/2}=300$ GeV and we take $\tan\beta=10$. Although $\mu>0$ has been used, the results for $\mu<0$ would be identical, due to the symmetry of the soft parameters under a shift of $\pi$ in $\theta$ and the fact that the results for the cross-section for $-\mu$, $M$, $A$ are equal to those with $\mu$, $-M$, $-A$.

All the points represented satisfy the experimental constraints on the lower masses of the supersymmetric particles and satisfy $m_h\ge91$ GeV. Small (green) dots represent points not fulfilling the \bsg\ constraint. Large dots do satisfy that constraint, and among these, blue points have $91$ GeV$\le m_h\le114$ GeV, while red dots satisfy the stronger lower bound for the Higgs mass $m_h>114$ GeV. Let us recall that this constraint on the Higgs mass holds in general for the cases with universal soft terms for $\tan\beta\lsim50$ and therefore it is the one we should consider here. However, due to the strong restrictions it imposes we prefer to show it explicitely.

As it could be expected, the experimental result for \asusy\ puts a strong lower bound on the masses of the supersymmetric particles for the cases with \asusy$<0$. For example, in the case shown in the figure, the accepted values are limited to those satisfying $m_{\tilde\chi_1^0}\gsim200$ GeV. For positive \asusy\ the constraint is still strong, comparable to that from \bsg, and becomes more restrictive for larger values of $\tan\beta$, but it is the lower bound on the lightest Higgs mass which puts a more severe constraint.

Let us now concentrate on those vacua with five-branes. These non-perturbative vacua also have generically $M\sim-A$. However their parameter space is much richer and we have found a more sophisticated pattern for the soft terms in the previous section. 
Let us first recall the parameter space of the theory in this case. 
Now in the expressions for the soft terms there are seven free parameters, which can be chosen as: the gravitino mass, $m_{3/2}$, two independent goldstino angles, $\theta$ and $\Theta$, the parameter $e_O$, the instanton number in the observable hyperplane, $\beta_O$, and the five-brane position and charge, $z$ and $\beta_1$, respectively.

Our analysis of \asusy\ will be carried out as follows. We will first fix the five-brane parameters ($z$, $\beta_1$) and the instanton number on the observable hyperplane ($\beta_O$). Using now Fig.\,\ref{regionsfig} and (\ref{constraintse}) we know the allowed values for the $e_O$ parameter. We will then fix $e_O$, requiring $V_O^{-1/6}$ to be close to $M_{GUT}$, and study specific examples with $e_O>0$ and $e_O<0$. Thus we are left with only four free parameters: $\tan\beta$, $m_{3/2}$, $\theta$, and $\Theta$. The values of $\tan\beta$, and $m_{3/2}$ will be fixed and the two goldstino angles varied over the whole range ($\theta\in[0,2\pi)$, $\Theta\in[0,\pi)$). The sign of the $\mu$ parameter is again irrelevant if we scan on this range of $\theta$, and $\Theta$, due to the existing symmetries in the soft terms.

\begin{figure}[!t]
\begin{center}
\epsfig{file=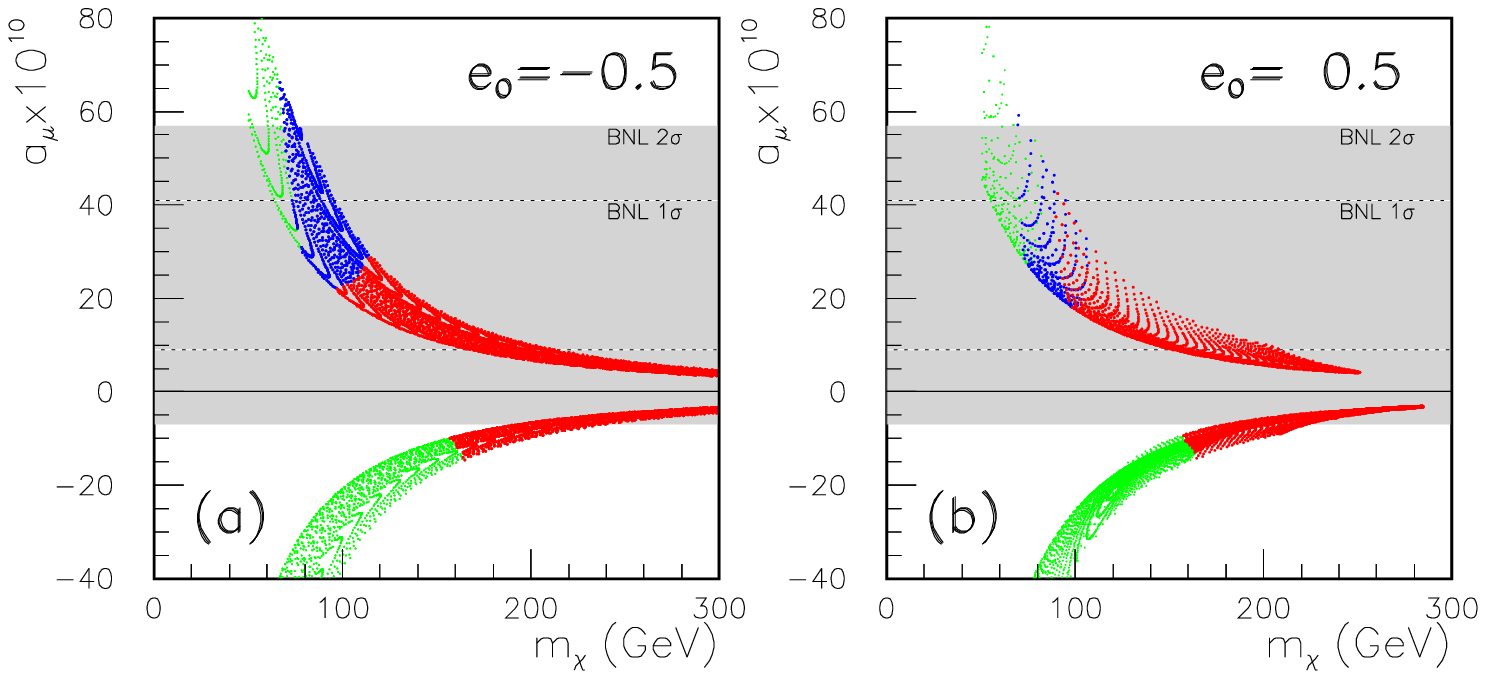}
\end{center}
\vspace*{-0.5cm}
\captions{a) Supersymmetric contribution to $\amu$ versus the neutralino mass in scenario (\ref{scenarioi}). b) The same for scenario (\ref{scenarioiii}). The colour convention of Fig.\,\ref{nofbg2.eps} is used.\label{mthg2.eps}}
\end{figure}

Let us first consider a case with negative $e_O$. We will choose (\ref{scenarioi}) as a representative example of this kind of scenarios. Accordingly to the restrictions on $e_O$, we will choose $e_O=-0.5$. The value of the GUT scale is in this case $V_O^{-1/6}\approx 2.8\times 10^{16}$ GeV, and therefore, the usual matter content of the MSSM suffices to obtain $\alpha_{GUT}\approx1/24$. We show in Fig.\,\ref{mthg2.eps}a the resulting \asusy\ versus the predicted neutralino mass for $\tan\beta=10$ and $m_{3/2}=300$ GeV. The \bsg\ constraint imposes a lower bound on the supersymmetric spectrum (in this case $m_{\tilde\chi^0}\gsim70$ GeV), but the most relevant constraint is that on the lightest Higgs mass, implying $m_{\tilde\chi^0}\gsim100$ GeV. Once all these constraints are applied, the maximum values for \asusy\ are obtained for small scalar masses $m\lsim 100$ GeV, which can even be as low as $60$ GeV. In this case $M\sim 270$ GeV, while $A\sim-160$ GeV. 

We will now exemplify the cases with positive values of $e_O$ with scenario (\ref{scenarioiii}), where $e_O=0.5$ will be used. As before, the value of the scale ($V_O^{-1/6}\approx 5.9\times 10^{16}$ GeV) is such that it is a good approximation to consider just the matter content of the MSSM in order to get coupling constant unification with $\alpha_{GUT}\approx1/24$. In this case, as shown in Fig.\,\ref{softratiose>0} gaugino masses lighter than scalar masses can be obtained. In order to enhance those regions, we take a larger value for the gravitino mass $m_{3/2}=500$ GeV. 
This case is depicted in Fig.\,\ref{mthg2.eps}b, where the same qualitative features as before are found. Now, after all the constraints are applied (in particular $m_h>114$ GeV is again the strongest one), the largest values of \asusy\ (\asusy$\lsim 40$) are obtained for $m\sim70$ GeV, $M\sim220$ GeV, and $A\sim-375$ GeV. Note that this case permits to obtain slightly larger values for \asusy and again due to the large (negative) values of $A$. In this case, there are few points with $m>|M|$, these would predict $20\lsim$\asusy$\lsim 30$, but in the end they are excluded by the \bsg\ constraint.

\begin{figure}[!t]
\begin{center}
\epsfig{file=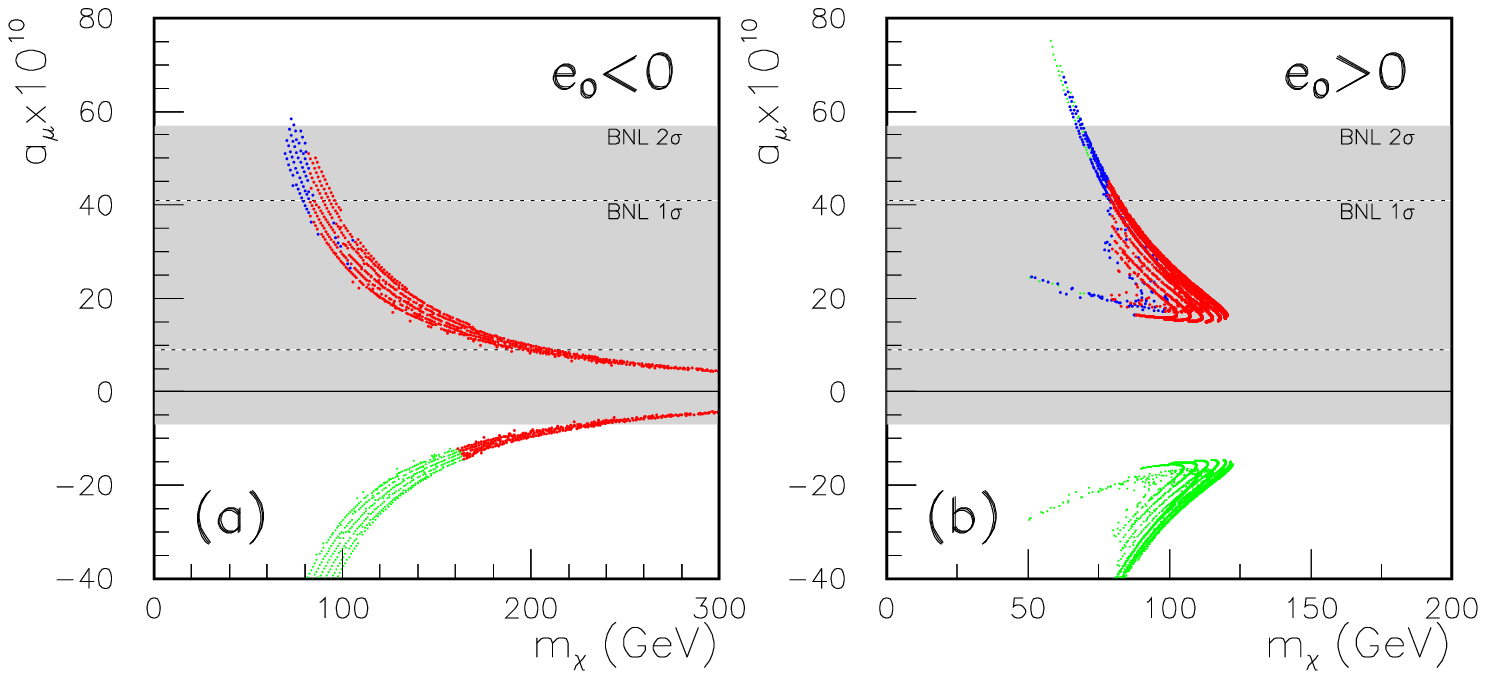}
\end{center}
\vspace*{-0.5cm}
\captions{a) Supersymmetric contribution to $\amu$ versus the neutralino mass when only the five-brane contributes to the breaking of supersymmetry for a generalization of scenario (\ref{scenarioi}), taking $-0.6<\sigma e_O<-0.1$ and the complete range for the five-brane positions. b) The same for a generalization of scenario (\ref{scenarioiii}), taking $\sigma e_O>0.1$. The colour convention of Fig.\,\ref{nofbg2.eps} is used.\label{fsft0g2.eps}}
\end{figure}

The two cases represented in Fig.\,\ref{mthg2.eps} for vacua with five-branes do not present therefore any qualitative difference with those scenarios without five-branes of Fig.\,\ref{nofbg2.eps}.

Let us finally illustrate the representative case where the breaking of supersymmetry is only due to the five-brane. We will work with the particular examples from the previous Section (see Figs.\,\ref{fsft0_iii} and \ref{fsft0_i}, where the corresponding soft terms are represented), for both signs of the $e_O$ parameter, corresponding to variations of scenarios (\ref{scenarioiii}) and (\ref{scenarioi}), respectively. The complete range of five-brane positions will be analyzed, as well as the values for the $e_O$ parameter. In order to obtain $V_O^{-1/6}\sim M_{GUT}$ we will only consider $-0.6<\sigma e_O<-0.1$, and $\sigma e_O>0.1$ (see Fig.\,\ref{scalesancient}). As we saw in Fig.\,\ref{scales-z}, a variation in the position of the five-brane does not induce a sizeable change in the value of the scales, and therefore no restriction on $z$ will be imposed. The results are shown in Fig.\,\ref{fsft0g2.eps}, following the same criteria explained above. The gravitino mass has been set to $m_{3/2}=300$ GeV. We find that these cases permit larger values of \asusy, up to $50$, and again this is due to $A$ being large and negative, while keeping $M$ relatively small. For example, the largest values for \asusy\ are found in Fig.\,\ref{fsft0g2.eps}a for $M\sim 200$ GeV, $m\sim 40$ GeV, and $A\sim -200$ (thus implying $A\sim -5m$). In Fig.\,\ref{fsft0g2.eps}b, we have $M\sim 200$ GeV, $m\sim 90 GeV$, and $A\sim -200$.
These results are slightly different to those of the vacua without five-branes and, in particular, the lower bound derived from this analysis in the neutralino mass also decreases, being now of the order of $75$ GeV.

Summarizing, the analysis of the supersymmetric contribution for $\amu$ has been performed for several representative examples of the parameter space of Heterotic M-theory. The results serve to determine the importance of the experimental constraint on \asusy\ in the different scenarios. In particular, we have seen that this is as compelling as \bsg\ and sets a lower bound on the supersymmetric spectrum. Although extensive regions in the parameter space can be found that fulfill these constraints, we will see in the following section that the fact that the lighter spectra are disfavoured will lead to small values of the neutralino-nucleon cross-section.


\section{Dark Matter}
\setcounter{equation}{0}
\label{sec_darkmatter}

Finally, using our former analysis of the soft terms, together with the results of the previous section, 
we can analyze how compatible is the parameter space of heterotic M-theory with the sensitivity of current dark matter detectors. 
Several analysis of dark matter in M-theory were performed in the standard embedding case \cite{bkl98-1,bkl98-5}, as well as in non-standard ones without five-branes \cite{klp00-1}, paying special attention to the calculation of the relic density. 
Dark matter implications of vacua with five-branes were investigated in the limit where the five-brane modulus is the only one responsible for the breaking of supersymmetry \cite{bkl01-1}.
However, the authors used previous soft-terms computed in the literature, where the corrections discussed along the present work were not included. 
In particular, the limit where the modulus of the five-brane is the only one responsible for supersymmetry-breaking turns out to be essentially different (see the discussion in Sec.~\ref{sec_softterms} and  Figs.\,\ref{fsft0_iii} and \ref{fsft0_i}). 
In this section we will briefly review the dark matter implications for both the standard and non-standard embeddings without five-branes, considering the recent experimental constraints discussed in the previous section. These will also be applied to the non-perturbative vacua with five-branes for which several representative examples will be analyzed. In particular, the five-brane dominance limit will be studied in detail.

The lightest neutralino, $\tilde\chi_1^0$, is one of the most promising candidates to solve the dark mater problem (see \cite{jkg96} for a review), being the lightest supersymmetric particle (LSP) in many models. However, as a weakly-interacting massive particle (WIMP), its typical scattering cross-section with a nucleon of a material inside a detector, $\crosssec$, is of the order of $10^{-8}$ pb. This is much below the sensitivity of current dark matter experiments (DAMA \cite{dama99,bcfs02-1}, CDMS \cite{cdms02-1}, UKDMC \cite{ukdmc98}, EDELWEISS \cite{edelweiss01-1}, IGEX \cite{canfranc00-1,igex01-1}, HEIDELBERG-MOSCOW \cite{heidelberg-moscow98-1}, HDMS \cite{hdms02}), which are sensitive to a $\crosssec$ of around $10^{-6}-10^{-5}$ pb. 
Thus if neutralinos were to be detected at the present experiments, we would have to find a mechanism which explains the enhancement in their typical interaction cross-sections of several orders of magnitude. Despite this fact, the DAMA collaboration reported a WIMP signal \cite{dama00} in their search for annual modulation (CDMS, IGEX, EDELWEISS and HDMS claim, however, to have excluded some of the regions in DAMA parameter space). 

There are several scenarios in the context of Supergravity (SUGRA) where such an enhancement may occur (see \cite{cgm02-1} for a recent review). It was pointed out in \cite{bdfs99-1,an99-1,aads00-1,cn00-1} that $\crosssec$ could reach the sensitivity region for the current dark matter detectors ($10^{-6}-10^{-5}$ pb) with non-universality of the soft supersymmetry-breaking terms. In particular, this is the case with nonuniversal scalar masses, for which $\crosssec\approx10^{-6}$ pb can be obtained for moderate values of $\tan\beta$ \cite{ad01-1,ads02-1}.
In addition, another possibility was pointed out in \cite{gkmt00-1}, where, inspired by superstrings, where the unification scale could be smaller than $10^{16}$ GeV, the sensitivity of the neutralino-nucleon cross-section to the value of the initial scale was studied. It was found that, the smaller the scale is, the larger the cross-section becomes. It was also argued that in the scenarios where the coupling constants do not unify the cross-sections are larger than in the case of gauge-coupling unification.

We do not have these possibilities in heterotic M-theory. The soft terms (\ref{escalares}-\ref{gaugino}) are universal, all the gauge couplings come from the same $E_8$ gauge group, and therefore, their value at the unification scale is the same, and, as we discussed in \cite{cm99-1}, and here in Sec.~\ref{sec_scales}, it is very unnatural to obtain low scales in this scenario. In fact, all the examples we will study here correspond to scales of the order of the phenomenologically accepted value for the GUT scale. 
Due to all these features, these could be considered as a subset into the parameter space of minimal Supergravity (mSUGRA), and in this sense, we do not expect to obtain high values for $\crosssec$. We will illustrate all this with specific examples.
We will be working with the usual formulas for the elastic scattering of relic LSPs on protons and neutrons that can be found in the literature (see \cite{jkg96} for a review). In particular, we will follow the re-evaluation of the rates carried out in \cite{efo00-1}, using their central values for the hadronic matrix elements. Our conventions for this section can be found in \cite{ckm01-1}.


\begin{figure}[!t]
\begin{center}
\epsfig{file=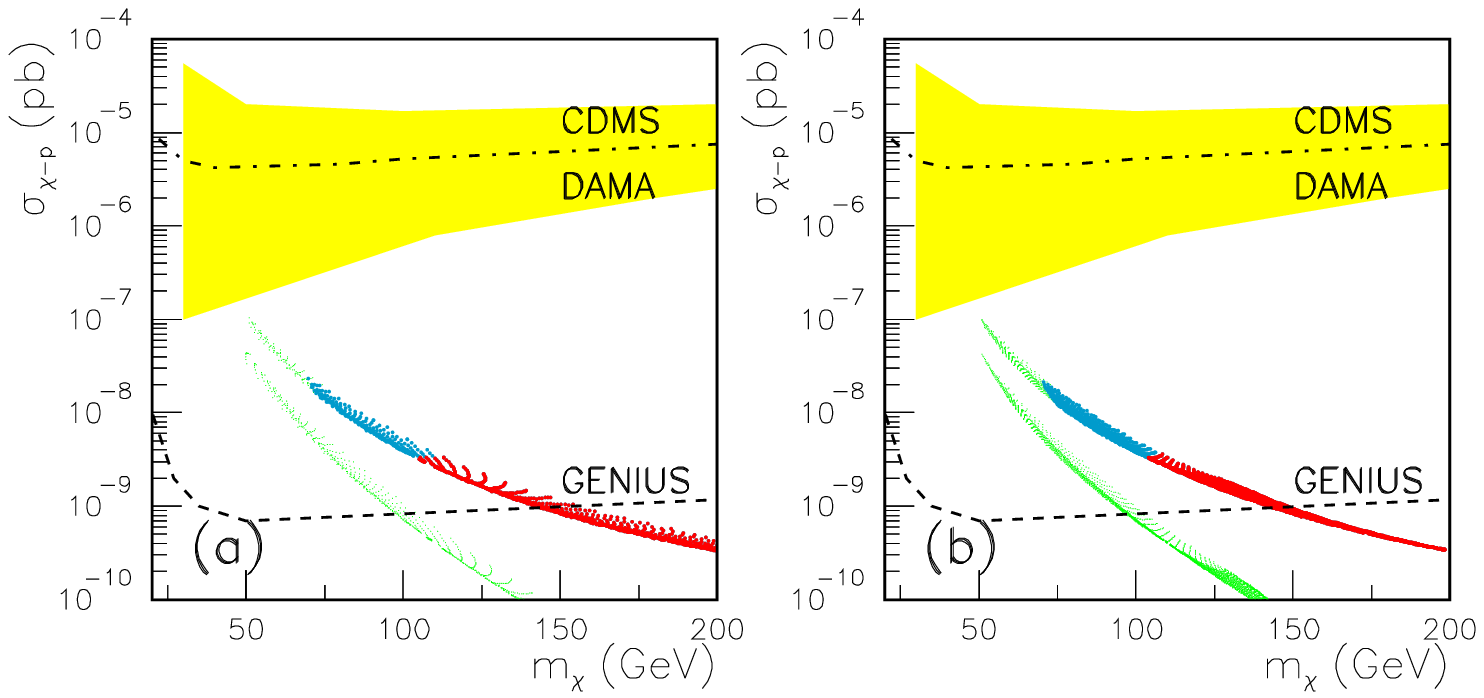}
\end{center}
\vspace*{-0.5cm}
\captions{a) Neutralino-nucleon cross-section versus neutralino mass for non-standard embedding without five-branes and  $-0.6\le\epsilon_O\le-0.1$. b) The same for the standard embedding and $0.1\le\epsilon_O<1$. The colour convention of Fig.\,\ref{nofbg2.eps} has been used. Current DAMA and CDMS limits and projected GENIUS limit are shown.
\label{nofbcross}}
\end{figure}

Let us first review the results for the standard and non-standard embedding cases without five-branes when the recent experimental constraints are take into account. As in the previous section, we choose $\beta_O=1$, and $\beta_O=-1$, respectively.
The values of the $\epsilon_O$ parameter will be chosen in order to guarantee $V_O^{-1/6}\sim M_{GUT}$ (i.e.,  $-0.6\le\epsilon_O\le-0.1$ and $0.1\le\epsilon_O<1$ for non-standard and standard embeddings, respectively). Again, we will take $m_{3/2}=300$ GeV, and $\tan\beta=10$, performing a variation of the goldstino angle, $\theta$, in $[0,2\pi)$.
Both cases are depicted in Fig.\,\ref{nofbcross}.
The experimental constraints put severe bounds, but
 still neutralinos as light as $\sim100$ GeV can be obtained (see Fig.\,\ref{nofbg2.eps}).
Once the lower limit on the Higgs mass is applied, the cross-section is as small as $\crosssec\sim3\times10^{-9}$ pb, far beyond the reach of present detectors, and only within the reach of the projected GENIUS. 
Although the predicted values for the cross-section increase, in principle, when larger values of $\tan\beta$ are taken into account, the experimental bounds become much more important in these cases (especially those corresponding to \bsg\ and \asusy), excluding larger regions in the parameter space and thus forbidding large values of $\crosssec$.

\begin{figure}[!t]
\begin{center}
\epsfig{file=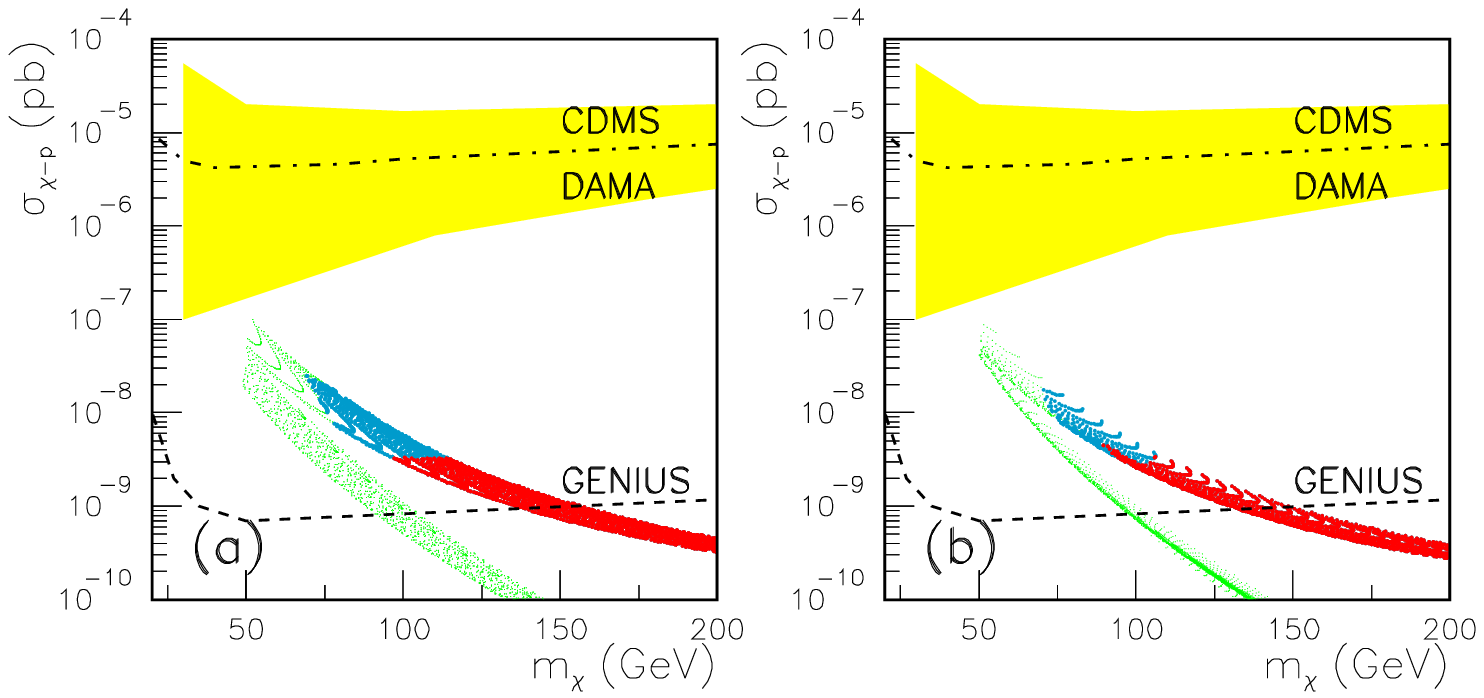}
\end{center}
\vspace*{-1cm}
\captions{a) Neutralino-nucleon cross-section versus neutralino mass for an example with five-branes and $e_O=-0.5$, corresponding to scenario (\ref{scenarioi}). b) The same, but for an example with $e_O=0.5$, corresponding to scenario (\ref{scenarioiii}). In both cases the goldstino angles have been varied over the whole range.
\label{fig_cross}}
\end{figure}

Let us now focus our attention to those vacua with five-branes.
Similarly to what we did in the previous section, we will concentrate on some representative examples of the parameter space. 
We will first analyze an example with $e_O<0$. In particular, we will consider again the scenario (\ref{scenarioi}), and we will fix $e_O=-0.5$, which, as we said in the previous section, corresponds to $V_O^{-1/6}\approx2.8\times10^{16}$ GeV.
In Fig.\,\ref{fig_cross} we show the resulting $\crosssec$ for $\tan\beta=10$, $m_{3/2}=300$ GeV, and a complete scan on the goldstino angles.
We can see how \bsg\ and \asusy\  bounds limit the value of the cross-section to $\crosssec\lsim2\times10^{-8}$ pb. 
The stronger of these constraints is \bsg, as we see from Fig.\,\ref{mthg2.eps} the values of \asusy\ remain within the $2\sigma$ error of the experimental value. These results do not differ from those for vacua without five-branes and the discussion above holds also in this case.
Lower bounds on the soft parameters can be derived after all the constraints have been applied, and in this case we obtain $M\gsim240$ GeV, $m\gsim50$ GeV, and $-A\gsim-160$.
Finally, although there was a region of the parameter space where $m/|M|>1$ (see Fig.\,\ref{softtermse<0}), the gaugino masses for this region are so small that these points have been excluded by experimental cuts. If we increase the gravitino mass, these points would eventually appear, but this would not imply any significant increase on the cross-section.

We can also study an example where $e_O>0$. We consider again the scenario (\ref{scenarioiii}) with $e_O=0.5$, predicting $V_O^{-1/6}\approx5.9\times10^{16}$ GeV.  
For the same values of $\tan\beta$ as before, and $m_{3/2}=500$ GeV in order to enhance those regions with $m>|M|$, we plot in Fig.\,\ref{fig_cross} the resulting neutralino-nucleon cross-section versus the neutralino mass. We do not find any significant difference with the case for negative $e_O$. 
Again, in this case, there were regions in the parameter space where the scalar masses were bigger than gaugino masses (see Fig.\,\ref{softtermse>0}), and once more, these regions are excluded by experimental bounds due to its low gaugino mass. As before, using larger values for the gravitino mass we can avoid these bounds for part of these regions, but this does not imply any increase of $\crosssec$.
In this case, the same strong constraints as before apply, leading to a small cross-section. The lower bounds on the soft parameters are now also very similar: $M\gsim220$ GeV, $m\gsim70$ GeV and $-A\gsim-230$ GeV.


Let us finally analyze the special case where the five-brane modulus is the only one responsible for supersymmetry-breaking. We will work on the same example of the previous section. As we mentioned before, the supersymmetric spectrum and the neutralino-nucleon cross-section with five-brane dominance were studied in \cite{bkl99-1} and \cite{bkl01-1}, respectively, but the new structure of the soft terms makes it necessary to revisit this case. The parameter space is now reduced, in the sense that the goldstino angles are now fixed for a particular choice of $e_O$, as in (\ref{fsft0thetas}). 
\begin{figure}
\begin{center}
\epsfig{file=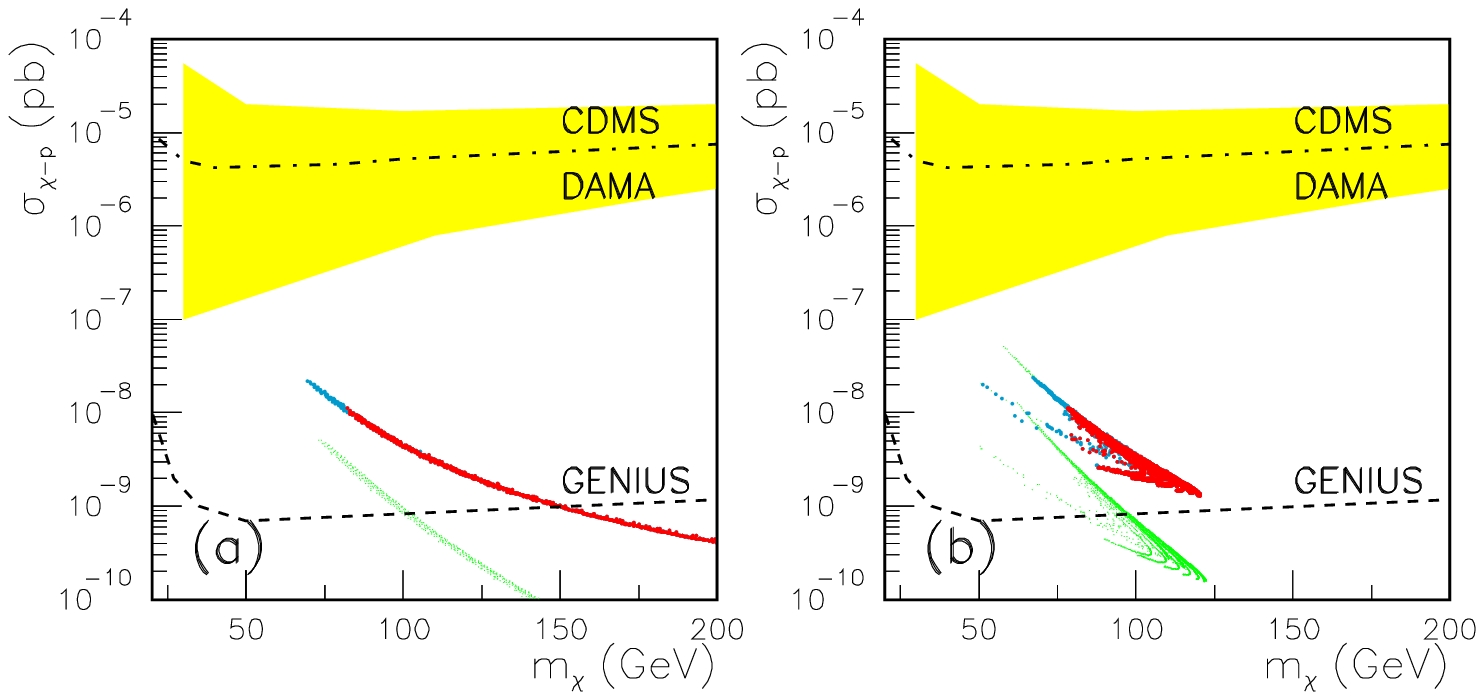}
\end{center}
\vspace*{-1cm}
\captions{a) Neutralino-nucleon cross-section for a case of five-brane dominance, for $-0.6<\sigma e_O<-0.1$ and for different positions of the five-brane $z$, covering the range $[0.1]$. b) The same but for $\sigma e_O> 0.1$.
\label{fsft0dark}}
\end{figure}

Let us first concentrate on the example or negative $e_O$ depicted in Fig.\,\ref{fsft0_i}. We will consider several values for the position of the five-brane, covering the whole range along the orbifold ($z\in [0,1]$) and vary $e_O$ along the allowed range (which depends on $z$). As we did in the analysis of \asusy, we will consider only those values satisfying $-0.6<\sigma e_O<-0.1$ in order to obtain $V^{-1/6}_O$ of the order of $M_{GUT}$ (see Fig.\,\ref{scalesancient}). The results are shown in Fig.\,\ref{fsft0dark}a. As in the previous examples, the gravitino mass is fixed at $m_{3/2}=300$ GeV and $\tan\beta=10$.
Now, the fact that scalar masses are much smaller than gaugino masses implies that for an important part of the parameter space the neutralino is not the LSP. In particular, all the region with $z\lesssim 0.7$ is excluded for this reason. 
Obviously, from Fig.\,\ref{fsft0_i}, larger values for the cross-section will be obtained for $e_O\sim-0,1$. This is indeed the case, and the soft parameters that produce $\crosssec\sim10^{-8}$ pb are: $m\sim50$ GeV, $M\sim200$ GeV, and $A\sim-180$ GeV.

For positive values of the parameter $e_O$ the results for the cross-section are very similar, as we show in Fig.\,\ref{fsft0dark}b. These results have been computed from the soft terms presented in Fig.\,\ref{fsft0_iii}, for diverse positions of the five-brane. Again, the values of $e_O$ have been restricted to those for which considering $M_{GUT}\sim 3\times 10^{16}$ GeV was a reasonable approximation. From Fig.\,\ref{scalesancient} it can be seen that this the case if $\sigma e_O\gsim 0.1$. 
Now, the larger values of the cross-section are obtained for values of $e_O$ close to $0.1$ (the upper limit we have imposed). As we see from Fig.\,\ref{fsft0_iii}, the spectrum is lighter in this case and therefore the cross-section increases. The higher values $\crosssec\sim10^{-8}$ pb are obtained for $m\sim 90$ GeV, $M\sim 195$ GeV, and $A\sim-215$ GeV.

Summarizing, as it could be expected, the predictions of heterotic M-theory for the neutralino-nucleon cross-section are too low to be probed by the present dark matter detectors. Only future experiments, as e.g. GENIUS, would be able to explore such low values. This is due to the universality of the soft terms and the fact that intermediate scales cannot be obtained in a natural way. 
In the scenarios with five-brane dominance in the breaking of supersymmetry, small values of $|e_O|$ are preferred in order to obtain larger $\crosssec$. 
Last, although $m>|M|$ was shown to be possible, this does not have an important influence on the cross-section predictions. 
In such cases, very high values for the gravitino mass are needed in order to satisfy the chargino bound.

\section{Conclusions}
\setcounter{equation}{0}
\label{sec_conclusions}

In the present paper we have analyzed several phenomenological aspects of heterotic M-theory with five-branes. Using recent results for the five-brane contribution to the \kah\ potential and gauge kinetic functions, and the correct identification of the five-brane modulus, we have performed a systematic analysis of the parameter space of the theory when one five-brane is introduced in the bulk, finding the restrictions on the parameter space that result from requiring the volume of the Calabi-Yau to remain positive. 

We have then concentrated on the evaluation of the different scales of the theory, namely the $11$-dimensional Planck scale, $M_{11}$, the compactification scale, associated to the volume of the Calabi-Yau, $V_O^{-1/6}$, and the orbifold scale $(\pi\rho)^{-1}$, finding very similar results to those obtained for vacua without five-branes. 
In particular, we have shown how the phenomenologically favoured value for the scale is easily recovered for most of the natural choices of the parameters. Also, intermediate scales have been shown to appear. This is the case of the limit $e_O\to-1$, which implies a hierarchy problem in the VEVs of the dilaton and modulus fields.
Although in most of the cases the dependence of the different scales on the five-brane position, $z$, is negligible, this is not the case for the particular choice $\beta_O=0,\ \beta_1=1$. For this setup we have found that if the five-brane is very close to the hidden fixed hyperplane, intermediate values for the scale can be obtained. However, the amount of fine-tuning in $z$ renders this possibility extremely unnatural and therefore has been discarded.

The soft supersymmetry-breaking terms have been computed for the effective theory described in the previous sections. 
The presence of the five-brane induces off-diagonal terms in the \kah\ metric and the analysis becomes more involved. In particular, the expressions for the parameterization of the $F$-terms in terms of the goldstino angles are now more complicated. 
We have found that a new pattern for the soft terms arises, due to the inclusion of a five-brane. In particular, scalar masses larger than gaugino masses are now more easily obtained for many natural choices of the parameters. 
We have analyzed this possibility in representative examples of the parameter space, investigating different limits on it. In this sense, the special case where the five brane, despite being present, does not contribute to the breaking of supersymmetry (this is, $F^Z=0$) has been analyzed, as well as the limit where the breaking is only due to the five-brane modulus (this is, $F^S,\ F^T=0$). In both cases, scalar masses larger than gaugino masses have been shown to appear, being this more easily fulfilled when the five-brane is close to the hidden sector hyperplane.

Using the results of the previous sections, we have derived the supersymmetric spectrum and computed the theoretical predictions for the supersymmetric contribution to the muon anomalous magnetic moment, \asusy. Asking for compatibility at the $2\sigma$ level with the recent experimental result leads to severe constraints on the parameter space. Again, we have analyzed the most representative cases of the parameter space.

Finally, including these constraints, together with the experimental constraints on the masses of the supersymmetric particles as well as those derived from the theoretical prediction of the \bsg\ branching ratio, we have computed the neutralino-nucleon cross-section in this construction. 
Due to the universality of the soft supersymmetry-breaking terms and the fact that the most natural value for the initial scale is of order $10^{16}$ GeV, the parameter space can be considered as a subset of mSUGRA.
Therefore, the predicted cross-section is very low, far beyond the reach of the present dark matter experiments.

\vspace{5ex}
{\Large{\bf Acknowledgements}}
\vspace{2ex}

 We would like to thank J. J. Manjar\'\i n for useful conversations and discussions. 
D.G.Cerde\~no acknowledges the financial support of the Comunidad de Madrid through a FPI grant. The work of C. Mu\~noz was supported in part by the Ministerio de Ciencia y Tecnolog\'\i a under contract FPA2000-0980, and the European Union under contract HPRN-CT-2000-00148.


\bibliography{bibbib}
\end{document}